\documentclass[aps, prl, twocolumn, reprint]{revtex4}

\usepackage{array}
\usepackage{graphicx}
\usepackage{dcolumn}
\usepackage{bm}
\usepackage{amsmath}
\usepackage{amsfonts}
\usepackage{amssymb}
\usepackage{amstext}   
\usepackage{amsthm}
\usepackage{dsfont}
\usepackage{color}
\usepackage{float}
\usepackage[utf8]{inputenc}
\usepackage[outdir=./]{epstopdf}
\usepackage{nameref}
\usepackage{subfloat}
\usepackage[colorlinks = true, citecolor = blue]{hyperref}
\usepackage[figure]{hypcap}
\usepackage[colorinlistoftodos]{todonotes}
\usepackage{mathtools}
\usepackage[writekey = false]{notes2bib}

\newcommand{\ket}[1]{\ensuremath{\left|#1\right\rangle}}

\newtheorem{theorem}{Theorem}

\DeclareMathOperator{\tr}{tr}

\begin{document}

\title{Classical simulation of quantum circuits by dynamical localization: \\
analytic results for Pauli-observable scrambling in time-dependent disorder}

\author{Adrian Chapman}
\email{akchapman@unm.edu}
\author{Akimasa Miyake}
\email{amiyake@unm.edu}
\affiliation{Center for Quantum Information and Control, Department of Physics and Astronomy, University of New Mexico, Albuquerque, NM 87131, USA} 

\date{\today}

\begin{abstract}
We extend the concept of Anderson localization, the confinement of quantum information in a spatially irregular potential, to quantum circuits. Considering matchgate circuits, generated by time-dependent spin-1/2 XY Hamiltonians, we give an analytic formula for the out-of-time-ordered correlator of a local observable, and show that it can be efficiently evaluated by a classical computer even when the explicit Heisenberg time evolution cannot. Because this quantity bounds the average error incurred by truncating the evolution to a spatially limited region, we demonstrate dynamical localization as a means for classically simulating quantum computation and give examples of localized phases under certain spatio-temporal disordered Hamiltonians. 
\end{abstract}

\maketitle

\emph{Introduction.---}A peculiar phenomenon exhibited uniquely by quantum lattice systems is the suppression of conductance in the presence of disorder. This effect, known as \emph{Anderson localization} \cite{anderson1958absence, lahini2010quantum, abrahams2010years, crespi2013anderson} in the single-particle setting and \emph{many-body localization} \cite{gornyi2005interacting, basko2006metal, oganesyan2007localization, pal2010manybody, vosk2014dynamical, lev2015absence, bera2015manybody} in the interacting-multi-particle regime, is a result of interference, which confines a local disturbance to a bounded region near its initial position for a very long time. As a result, these systems do not act as thermal reservoirs for their own subsystems \cite{deutsch1991quantum, pal2012manybody, nandkishore2015manybody, huse2013localization, kjall2014manybody}, since local subsystems retain information about their initial conditions forever. 

An important consequence of the fact that local quantum information does not mix, or scramble, among nonlocal degrees of freedom in localizing systems is that many properties of these systems can be efficiently simulated classically. Such properties include local integrals of motion \cite{kim2014local, xu2015entanglement}, Hamiltonians' eigenbases \cite{bauer2013area, xu2015entanglement, eisert2015eigenvectors, huang2015efficient, yu2015mps, cirac2016efficient}, unitary time-evolution operators \cite{burrell2007bounds, znidaric2008manybody}, and samples from their output distributions \cite{deshpande2017complexity}. In light of these many results, we 
ask the question of whether localization could manifest in time-dependent quantum systems, such as those performing a quantum computation. If this is possible, then it would allow for the efficient simulation of an otherwise apparently complex quantum algorithm by classical means. 

However, there are currently very few prior investigations into localization in the time-dependent regime. Initial explorations into fluctuating disorder \cite{osborne2009information, sacha2016anderson} and Floquet circuit ensembles \cite{chandran2015semiclassical} suggest that a form of localization persists in these time-dependent cases, yet few analytic results are known in general. In this work, we consider the setting of nearest-neighbor matchgate circuits, which are generated by time-dependent spin-1/2 XY Hamiltonians, and can be mapped onto the dynamics of free-fermions with arbitrarily time-dependent single-particle Hamiltonians by the Jordan-Wigner transformation \cite{knill2001fermionic, terhal2002classical, bravyi2002fermionic, bravyi2006universal, jozsa2008matchgates, melo2013power, brod2014computational}. These circuits therefore constitute the natural framework in which to study the generalization of Anderson localization to quantum circuits. 

Despite the encoding by free-fermion dynamics, some properties of these circuits are not known to be classically simulable. The example we consider is the Pauli-expectation value $\langle U^{\dagger} X_j U \rangle$ on an arbitrary qubit $j$ in the output of a matchgate circuit $U$ from an initial arbitrary product state. Despite being local in the qubit picture, this observable takes the form of a long-range correlation function in the fermion picture and requires exponential resources to simulate by brute-force. We solve this problem for circuits $U$ describing localizing dynamics by exploiting the confinement of their measurement observables in the Heisenberg picture. For time-independent Hamiltonians, this confinement is described formally by the so-called \emph{zero-velocity Lieb-Robinson bound} \cite{hamza2012dynamical}

\begin{align}
||[A, B(t)]|| \lesssim \min{(|t|, 1)} e^{-\eta d(A, B)}.
\label{eq:ZVLRBound}
\end{align} 

\noindent This inequality states that the degree of noncommutativity between the local observable $A$ and time-evolved observable $B(t) \equiv U^{\dagger} B U$ initially separated by lattice distance $d(A, B) > 0$, such that $[A, B] = 0$, is exponentially decaying with decay constant $\eta > 0$. It gives an effective speed at which disturbances propagate \cite{lieb1972finite, nachtergaele2005liebrobinson, hastings2006spectral, nachtergaele2010liebrobinson}, which goes to zero with increasing propagation time in localizing systems. Correlations between distant lattice sites take exponential time to develop \cite{bravyi2006liebrobinson}.

In the case where $A$ and $B$ are unitary, and the norm taken is the Frobenius norm $||O||^2 \equiv \tr{\left(O^{\dagger} O \right)}$, the left-hand side of (\ref{eq:ZVLRBound}) is known as the infinite-temperature out-of-time-ordered correlation function (OTO correlator). This quantity has arisen as a useful diagnostic tool for studying scrambling in chaotic quantum chaotic systems \cite{maldacena2015bound, yoshida2016chaos, aleiner2016microscopic, halpern2017fluctuation, ding2017otooperator}, including black holes \cite{shenker2014black, kitaev2014hidden, kitaev2015simple, roberts2015localized}, and recently, for many-body localization \cite{chowdhury2016slow, chen2016universal, huang2016oto, he2016characterizing, fan2016oto}. As a first result, we provide an analytic formula for this quantity when $A$ and $B$ are Pauli observables, and the time evolution is described by a matchgate circuit. This is surprising considering that the evolution itself cannot even be stored efficiently by a classical computer in general, and so it constitutes an exponential speedup over the brute-force method. We next show that this quantity bounds the average-case change in expectation-value magnitude from truncating the Heisenberg evolution of $B(t)$ to a subset of qubits and thus provides a measure of the expected error incurred by such truncation. Finally, we provide numerical analysis verifying the bound (\ref{eq:ZVLRBound}) for two natural models of time-dependent disorder and construct phase diagrams demonstrating their transitions to localizing dynamics and subsequent classical simulability.

\emph{Background.---}Define a matchgate $G(V, W)$ to be the following 2-qubit unitary, written in the (ordered) computational basis $\{\ket{00}, \ket{01}, \ket{10}, \ket{11}\}$ as

\begin{align}
G(V, W) = \begin{pmatrix}
V_{00} & 0 & 0 & V_{01} \\
0 & W_{00} & W_{01} & 0 \\
0 & W_{10} & W_{11} & 0 \\
V_{10} & 0 & 0 & V_{11}
\end{pmatrix} \rm{,}
\end{align}

\noindent where $V, W \in \mathbf{SU}(2)$ are single-qubit unitaries (crucially, $\det V = \det W$). $G(V, W)$ preserves the eigenspaces of $Z \otimes Z$ and so may be written as $e^{i \mathcal{L}}$, where $\mathcal{L}$ is an element of the vector space spanned by $\{X \otimes X, X \otimes Y, Y \otimes X, \\ Y \otimes Y, Z \otimes I, I \otimes Z \}$. When the 2 qubits on which $G(V, W)$ acts are a nearest-neighboring pair, $\mathcal{L}$ is an instance of the 2-qubit spin-1/2 XY model. Such a Hamiltonian on $n$ qubits is quadratic in the $2n$ Majorana operators $\{c_{\mu}\}_{\mu = 1}^{2n}$, given by the Jordan-Wigner transformation

\begin{align*}
c_{2k - 1} \equiv \otimes_{j = 1}^{k - 1} Z_j \otimes X_k \mbox{\hspace{10mm}} c_{2k} \equiv \otimes_{j = 1}^{k - 1} Z_j \otimes Y_k \mathrm{,}
\end{align*}

\noindent where $c_{\mu}^2 = I$ and $c_{\mu} c_{\nu} = -c_{\nu} c_{\mu}$ for all $\mu \neq \nu \in \{1, \dots, 2n \}$. It is straightforward to verify that any such unitaries $U$ generated by quadratics in the Majorana modes form a group, and that this group is exactly that of circuits composed of nearest-neighbor matchgates \cite{jozsa2008matchgates}. Furthermore, such $U$ preserve the number of Majorana modes, as

\begin{align}
U^{\dagger} c_{\mu} U = \sum_{\nu = 1}^{2n} u_{\mu \nu} c_{\nu} \rm{,}
\label{eq:modepreserve}
\end{align}

\noindent where $\mathbf{u} \in \mathbf{SO}(2n)$ is a $2n \times 2n$ orthogonal matrix. We introduce a \emph{Majorana configuration} as an ordered tuple of indices $\vec{\alpha} \equiv (\alpha_1, \alpha_2, \dots, \alpha_k)$ of degree $|\vec{\alpha}| \equiv k$ with $\alpha_j \in \{1, \dots, 2n \}$ and $\alpha_j < \alpha_{j + 1}$ for all $j \in \{1, \dots, k \}$. The corresponding \emph{Majorana configuration operator} is the ordered product $C_{\vec{\alpha}} \equiv \prod_{j = 1}^{|\vec{\alpha}|} c_{\alpha_j}$, with Majorana indices ascending from left to right. Finally, denote by $\mathbf{u}_{\vec{\alpha} \vec{\beta}}$ the submatrix of $\mathbf{u}$ given by taking the rows indexed by $\vec{\alpha}$ and the columns indexed by $\vec{\beta}$, i.e. $(\mathbf{u}_{\vec{\alpha} \vec{\beta}})_{jk} \equiv u_{\alpha_j \beta_k}$.

Majorana configuration operators transform under matchgate evolution as (see Appendix~\hyperref[sec:CCAmps]{A})

\begin{align}
U^{\dagger} C_{\vec{\alpha}} U & = \sum_{\{\vec{\beta} | |\vec{\beta}| = |\vec{\alpha}| \}} \det \left( \mathbf{u}_{\vec{\alpha} \vec{\beta}} \right) C_{\vec{\beta}}  \rm{.}
\label{eq:configev}
\end{align} 

\noindent That is, the degree of a Majorana configuration operator is preserved, and configuration transition amplitudes are given by determinants of the corresponding single-mode transition submatrices. We also note that

\begin{align}
Z_k = -i C_{\left(2k - 1, 2k\right)} \mbox{\hspace{2mm}} \mathrm{and} \mbox{\hspace{2mm}} X_k = (-i)^{k - 1} C_{\left(1, \dots, 2k - 1 \right)} \rm{.}
\end{align}

\noindent From Eq.~(\ref{eq:configev}), we see that the Heisenberg evolution of $Z_k$ will always consist of ${{n}\choose{2}}$ terms, regardless of $k$. However, that of $X_k$ will consist of ${{n}\choose{2 k - 1}}$ terms, which may scale exponentially with $n$ if $k$ also scales with $n$, such as for $X_{\lfloor n/2 \rfloor}$ in the center of the chain. This is reflected in the fact that $\langle U^{\dagger} Z_k U \rangle$ can always be computed efficiently by a classical computer when the expectation values $\langle C_{\vec{\beta}} \rangle$, for $|\vec{\beta}| = 2$, can be, such as for product input \cite{jozsa2008matchgates}. On the other hand, $U^{\dagger} X_k U$ cannot even be stored efficiently on a classical computer in the worst case, so the same strategy will not work \bibnote{Though the distribution of such a measurement can be sampled efficiently, a weaker form of simulation, by the method in Ref.~\cite{brod2016efficient}}. Nevertheless, localization will provide a means to efficiently approximate this quantity, as we state formally below.

\emph{Analytic results.---}We are able to efficiently calculate the left-hand side of (\ref{eq:ZVLRBound}) in our setting by our first result

\begin{theorem}
  \label{thm:efficientoto} (Analytic OTO correlator)
The OTO correlator for Pauli observables $A \equiv i^a C_{\vec{\eta}}$ and $B \equiv i^b C_{\vec{\alpha}}$ may be computed analytically as

\begin{align*}
\frac{1}{2^{n + 2}}||[A, U^{\dagger}BU]||^2 = \frac{1}{2} \left\{1 \pm \det [\mathbf{u}_{\vec{\alpha} [2n]} (\mathbf{I} - 2\mathbf{P}_{\vec{\eta}}) \mathbf{u}^{\mathrm{T}}_{[2n] \vec{\alpha}}] \right\} \rm{,} 
\end{align*}

\noindent where $(\mathbf{P}_{\vec{\eta}})_{jk} = 0$ if $j \notin \vec{\eta}$ or $k \notin \vec{\eta}$, and $(\mathbf{P}_{\vec{\eta}})_{jk} = \delta_{jk}$ otherwise (i.e. $\mathbf{P}_{\vec{\eta}}$ is the projector onto the modes $\vec{\eta}$). The sign factor is simply $(-1)^{|\vec{\alpha}||\vec{\eta}| + 1}$.
\end{theorem}

\noindent The result follows from the Cauchy-Binet formula (see Appendix~\hyperref[sec:ProofThm1]{B}). In fact, it is possible to modify the Cauchy-Binet formula to obtain an analyltic calculation of the OTO correlator when $A \equiv \bm{n}_s \cdot \bm{\sigma}_s$ is any single-site Pauli observable (Appendices~\hyperref[sec:MCBForm]{C} - \hyperref[sec:ExactOTO]{E}). This allows us to regard the OTO correlator as a quadratic form $\mathbf{M}_s$, as
	
\begin{align}
\frac{1}{2^{n + 2}}||[\mathbf{n}_s \cdot \bm{\sigma}_s, U^{\dagger}BU]||^2 \equiv \mathbf{n}^{*}_s \cdot \mathbf{M}_s \cdot \mathbf{n}_s
\label{eq:quadform}
\end{align} 

When the bound (\ref{eq:ZVLRBound}) holds, we can efficiently approximate the evolution in Eq.~(\ref{eq:configev}) by truncating the sum to those $\vec{\beta}$ whose support lies strictly within a constant subset of qubits. We model this truncation by the action of a completely depolarizing channel

\begin{align}
\mathcal{E}_s(O) = \frac{1}{4} \left(O + \sum_{k \in \{x, y, z\}}  \sigma_s^k O \sigma_s^k\right) \mathrm{,}
\label{eq:depolardef}
\end{align}

\noindent which takes any single-qubit operator to its identity component. With our second result, we show that this truncation incurs a bounded error in the average case (see Appendix~\hyperref[sec:thm2proof]{F}):

 \begin{theorem}
\label{thm:otobound} (Average disturbance by truncation)
Let $\mathcal{E}_s$ be the completely depolarizing channel on qubit $s$. The average change in expectation-value magnitude of $U^{\dagger}BU$ under depolarization on a set of qubits $S$ is bounded by the OTO correlator as

\begin{align*}
\overline{|\langle U^{\dagger}BU \rangle - \langle (\otimes_{s \in S} \mathcal{E}_s)[U^{\dagger}BU] \rangle|} \leq \sum_{s \in S}\sqrt{\mathbf{n}^{*}_{s} \cdot \mathbf{M}_s \cdot \mathbf{n}_{s}} \rm{,}
\end{align*} 

\noindent where $\overline{(\cdot)}$ denotes an average over a product basis whose Bloch axes are orthogonal to the vectors $\{\mathbf{n}_s\}_{s}$, and $\mathbf{M}_s$ is as defined in Theorem \ref{thm:efficientoto}.
\end{theorem}

\emph{Numerical example.---}Theorem \ref{thm:efficientoto} is valid for every unitary time-evolution satisfying (\ref{eq:modepreserve}). However, we will narrow the focus of our numerical analysis to two specific Hamiltonian models, each of the form:

\begin{align}
H(t) = \sum^{n - 1}_{j = 1} 2 \mu_j (t) \left(X_j X_{j + 1} + Y_j Y_{j + 1}\right) + \sum^{n}_{j = 1} 2\nu_j (t) Z_j
\label{eq:parentham}
\end{align}

\noindent In Model 1, we allow the local disorder to fluctuate in time about some mean static disorder, and in Model 2, we allow interactions to fluctuate in space and time about mean translationally invariant interactions. The flucutations are chosen as independent, identically distributed random samples taken from the interval $[-\Delta, \Delta]$ every period $\delta t = 0.25$. We vary the strength of the mean value and the fluctuation strength $\Delta$ for each model, keeping the remaining parameter fixed. This is intended to resemble a discrete-time control setup, wherein some of the parameters are constrained but others may be varied with some control strength. The static limit, for which $\Delta = 0$, is well-understood (see e.g. Ref.~\cite{burrell2007bounds}) and will provide a convenient reference point. These models are summarized in Table \ref{tab1}.

\begin{figure}
\includegraphics[width=\columnwidth]{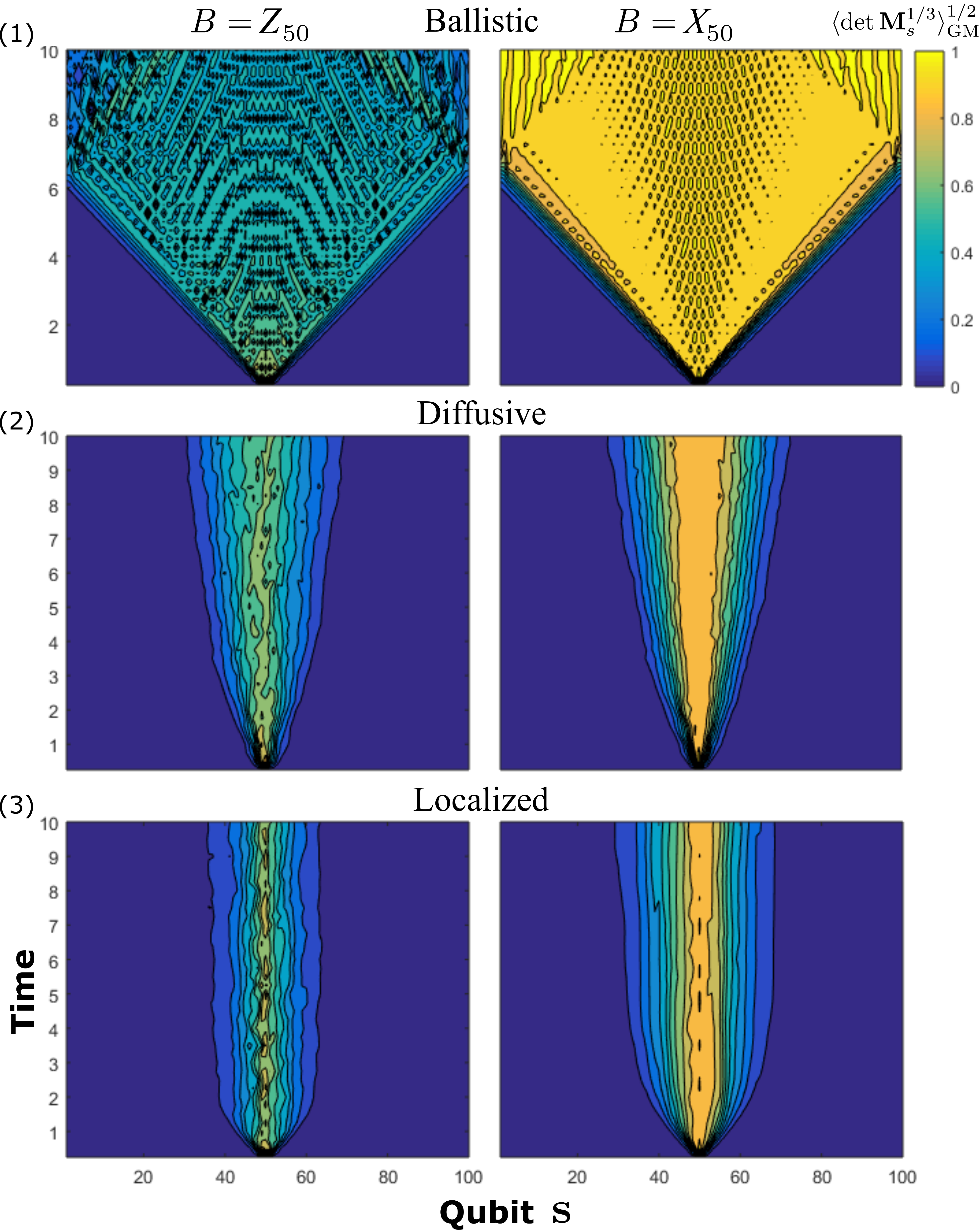}
\caption{(Color online) Typical light cones for $Z_{50}$ (left) and $X_{50}$ (right) propagating through the time-dependent disordered $XY$ model (Model 1) with mean disorder strength $\nu$ and fluctuation strength $\Delta$, for $n = 100$ qubits. Light cones are taken as the geometric mean of the determinant over 10 disorder realizations to preserve exponential decay. Representatives of the ballistic ($\nu = 0$, $\Delta = 0$, top), diffusive ($\nu = 0.75$, $\Delta = 1$, middle), and localized ($\nu = 2$, $\Delta = 0$, bottom) phases are shown. We note that, while the light cone envelopes for $Z_{50}$ and $X_{50}$ are always nearly the same, the interior of that for $X_{50}$ generally has a higher value.}
\label{fig:propagation}
\end{figure}

As the Hamiltonian (\ref{eq:parentham}) is quadratic in the Majorana operators, its time-evolution operator may be expressed as a matchgate circuit \cite{jozsa2008matchgates} and so may apply our Theorem \ref{thm:efficientoto}. In Fig.~\ref{fig:propagation}, we plot representative profiles of the OTO correlator (hereafter referred to as ``light cones") for $B = Z_{50}$ (left) and $B = X_{50}$ (right) for $n = 100$ in Model 1 in its ballistic (top), diffusive (middle), and localized (bottom) phases. In the localized propagation, for which static disorder $\nu = 2$ and $\Delta = 0$, the bound (\ref{eq:ZVLRBound}) is satisfied, and the observable support remains confined. As we increase fluctuations relative to static disorder in the middle plots, for which $\nu = 0.75$ and $\Delta = 1$, we see that time-dependent fluctuations induce a transition to diffusive propagation.

We identify the propagation phase of each profile by the taking principal singular component of its light cone, treated as a numerical matrix (see Appendix~\hyperref[sec:numanalysis]{G}). We argue that this gives an operationally meaningful, robust, and numerically inexpensive means of extracting the envelope and decay profile.

\begin{table}
\begin{tabular}{| c c | c | c | c  c|}
 \hline
 & \multicolumn{4}{l}{Parent Hamiltonian (\ref{eq:parentham})} & \\
 &\multicolumn{4}{l}{}& \\
 & \multicolumn{4}{l}{$H(t) = \sum^{n - 1}_{j = 1} 2 \mu_j (t) \left(X_j X_{j + 1} + Y_j Y_{j + 1}\right) + \sum^{n}_{j = 1} 2\nu_j (t) Z_j$} & \\
 &\multicolumn{4}{l}{}& \\
 \hline
& Model & Fluctuation & Fixed & Varying & \\
 \hline
& 1 & $\nu_j(t) = \nu_j + (2 / \delta t) \kappa_j(t)$ & $\mu = 1$ &  $\nu$, $\Delta$ & \\
& 2 & $\mu_j(t) = \mu + (2 / \delta t) \kappa_j(t)$ & $\nu = 1$ &  $\mu$, $\Delta$ & \\ \hline
\end{tabular}
\caption{Summary of the time-dependent models considered in this work, corresponding to the numerical phase diagram shown in Fig.~\ref{fig:phases}. $\nu_j \sim [-\nu, \nu]$ are chosen uniformly randomly, and the $\kappa_j(t) \sim [-\Delta, \Delta]$ are uniformly randomly sampled every $\delta t = 0.25$}.
\label{tab1}
\end{table}

\begin{figure}
\includegraphics[width=\columnwidth]{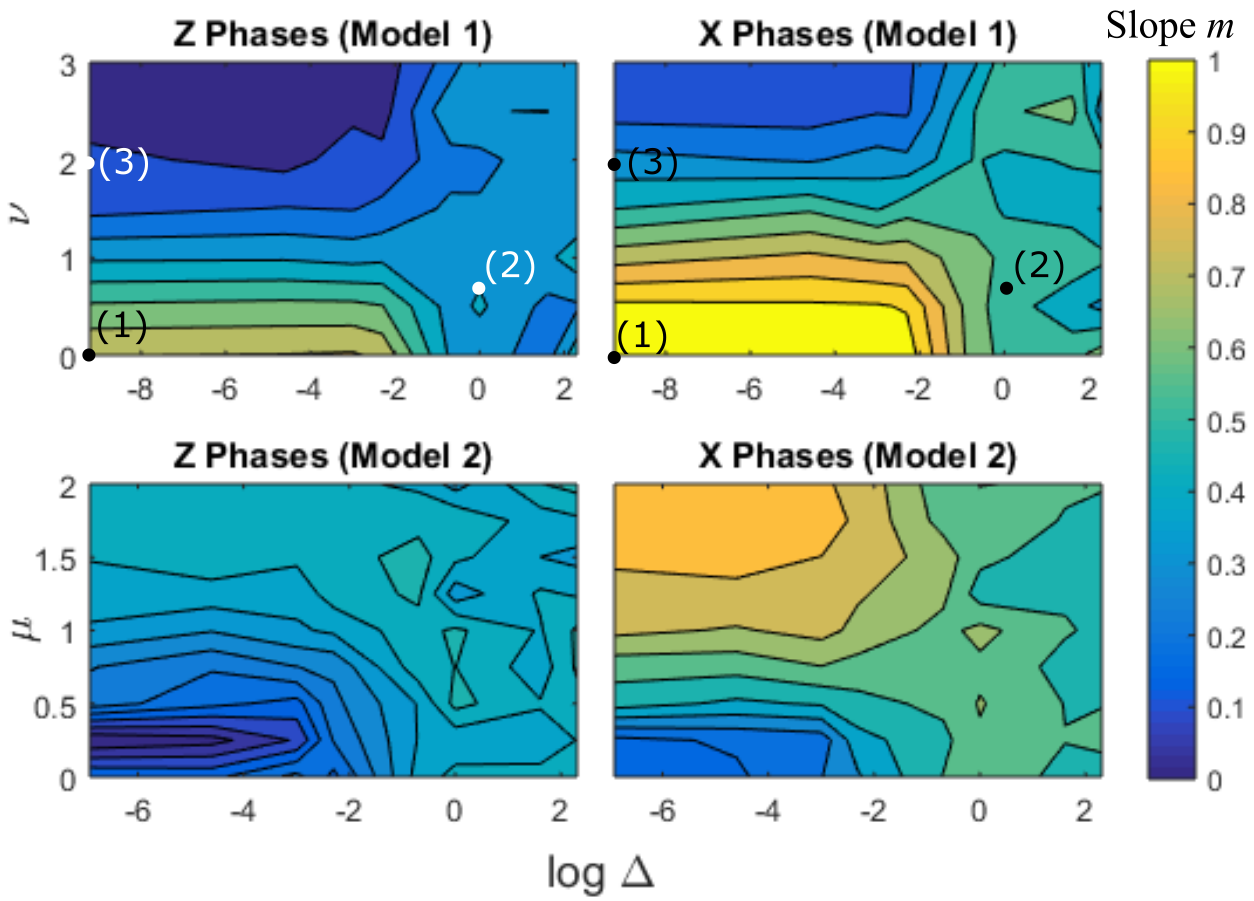}
\caption{(Color online) Phase diagrams for the propagation of $Z$ (left) and $X$ (right) in the presence of locally fluctuating disorder (top) and fluctuating interactions (bottom). These are given by taking the fitted slopes $m$ of the light cone envelopes. Though this parameter is continuous, we see that several natural regions emerge: ballistic ($m \approx 1$), diffusive ($m \approx 0.5$), and localized ($m \approx 0$). Numbered points correspond to the corresponding light cones in Fig. \ref{fig:propagation}, where points (1) and (3) on the \emph{y}-axis meant to be located at $\Delta = 0$. The extended localized phase demonstrates that localization survives under weak fluctuations.}
\label{fig:phases}
\end{figure}

We characterize the propagation phase by fitting the principal temporal component of each lightcone to a polynomial and extract the exponent of the leading-order term $t^m$. In Fig.~\ref{fig:phases}, we plot $m$ for $Z_{50}$ and $X_{50}$ for our two models for $n = 100$, as in Fig.~\ref{fig:propagation}, as phase diagrams. We identify the  \emph{ballistic} phase with regions where $m$ is very nearly one, the \emph{localized} phase with regions where $m$ is very nearly zero, and the \emph{diffusive} phase with regions where $m$ is nearly $0.5$.  With this identification, we see that as $\Delta \rightarrow 0$, our results agree with the known limit of static local disorder in Model 1. Similarly, as $\Delta$ becomes large, we see the emergence of a diffusive phase, which is consistent with the results put forth in \cite{osborne2009information}.
Finally, we see that, for small $\Delta \neq 0$, the localized phase survives. This indicates the existence of new matchgate circuits for which localization may be applied to classically simulate $\langle U^{\dagger} X_k U\rangle$ for arbitrary $k$ in a general product state input.

\emph{Discussion.---}We have presented examples where localization may be applied as a tool for classically simulating quantum circuits which were a priori believed to be classically intractable. This is achieved by an analytic calculation of the OTO correlator (presented in Appendix~\hyperref[sec:ExactOTO]{E}), followed by truncation to a subset of qubits for which this quantity falls below a certain threshold. 

One advantage of our method is that it gives the \emph{interior} of the light cone in addition to its envelope. In each phase, we see that the light cone interior for $X_{50}$ generally has a higher value than that for $Z_{50}$. This is a consistent difference between the profiles of these operators, which may have important consequences for the complexity required to exactly simulate their expectation values, in a similar fashion with \cite{yoshida2016chaos}. We attribute the emergence of a near-ballistic region in the $X$ phase diagram of Model 2, which is absent from the $Z$ diagram, to this observation. Though some amplitude propagates ballistically for both observables in this region, this only manifests as a spreading of the exponential tails for $Z$. For $X$ however, this is exhibited as a ballistic spreading of the high-amplitude region due to interference between its many constituent Majorana operators. This indicates that, at least in the presence of fluctuating interactions, propagation behavior between different local operators can be strikingly different. 

We empirically observe, however, that the difference between $X$ and $Z$ observables only emerges at late times, in the saturation value of the OTO correlator for each of these operators. By examining the constant-position slices from the light cones in Fig.~\ref{fig:propagation} as functions of time, we observe a characteristic exponential early-time behavior for these values, which is identical between $X$ and $Z$ propagation. Only as the growth of these operators saturate to roughly their constant values do the differences emerge. This suggests that the evolution of low-degree Majorana configurations may be useful as a good heuristic to observe the lightcone envelope, in a similar spirit to the treatment given in Ref.~\cite{swingle2018mpo} for finding a low bond-dimension matrix product operator approximation to the evolution of such observables in interacting-fermion systems.

Although we chose here so-called matchgate circuits, related to the time evolution of free fermions, because of their correspondence to Anderson localization, our method is expected to have further applications and extensions. On the former, for example, one may apply it to other random-circuit ensembles, such as those with Haar random matchgates, to study scrambling in general. A preliminary analysis indicates that propagation in this case seems to scale logarithmically, rather than polynomially. On the latter, it may be feasible to extend our method to analyze universal quantum computation by considering matchgate circuits acting on certain entangled input states, such as those of Ref.~\cite{bravyi2006universal}, as our method is independent of the input. In a similar way as Anderson localization has been extended to many-body localization, certain perturbative analysis (in analogy to that performed in \cite{bravyi2016improved} for Clifford circuits) could probe possible dynamical localization in general quantum circuits as well as simulating interacting fermions. We report some progress on this direction in an upcoming work.

This work was supported in part by National Science Foundation grants PHY-1521016.

\emph{Note Added:} Since the original submission of this work, the authors became aware of several new related works relating to scrambling in random quantum circuits \cite{haah2017operator}, matrix-product appproximations to the OTO correlator in the mixed-field quantum Ising chain \cite{swingle2018mpo}, and exact calculation of the OTO correlator for the transverse-field quantum Ising chain \cite{motrunich2018oto}. We expect that ours will contribute alongside these insightful papers to develop a unified picture of this growing field.

\onecolumngrid
\newpage
\begin{center}
\textbf{\large Supplementary Material for ``Classical simulation of quantum circuits by dynamical localization:
analytic results for Pauli-observable scrambling in time-dependent disorder''}
\end{center}

\setcounter{equation}{0}
\renewcommand{\theequation}{S\arabic{equation}}
\newcounter{Hequation}
\makeatletter
\g@addto@macro\equation{\stepcounter{Hequation}}
\makeatother

\section{Outline of Supplementary Material}

In the following sections, we prove Theorems \ref{thm:efficientoto} and \ref{thm:otobound} in the main text. In the first section, we show that the transition amplitudes between Majorana configuration operators -- ordered products of Majorana operators -- under matchgate evolution are given by determinants of submatrices of the single-Majorana transition matrix. We use this to simply prove Theorem \ref{thm:efficientoto} in the main text. Next, we show a modification to the Cauchy-Binet formula to compute sums of only those configuration transition amplitudes which involve a fixed background of Majorana operators. In Appendix \hyperref[sec:FPSum]{D}, we apply this formula to compute a sum of only those configuration transition amplitudes which involve a fixed parity on a given subset. Finally, we use the results of Appendices \hyperref[sec:CCAmps]{A}-\hyperref[sec:FPSum]{D} to arrive at our analytic calculation of the infinite-temperature OTO correlator with respect to any single-qubit observable, Eq.~(\ref{eq:quadform}) in the main text, in Appendix \hyperref[sec:ExactOTO]{E}. In Appendix \hyperref[sec:thm2proof]{F}, we prove a bound on the average-case change in expectation value induced by the depolarizing channel in terms of this quantity (Theorem \ref{thm:otobound}). Finally, we elaborate on some of our numerical techniques in Appendix \hyperref[sec:numanalysis]{G}.

\section{Summary of Notation}

\begin{table}[htbp]
\begin{center}
\begin{tabular}{l c p{15cm} }
$\vec{\alpha}$ & $\equiv$ & $\left(\alpha_1, \alpha_2, \dots, \alpha_k \right)$, where $\alpha_j < \alpha_{j + 1}$, for all $\alpha_j \in \{1, 2, \dots, N\}$ for some $N \in \mathds{N}$, and $j \in \{1, 2, \dots, k - 1\}$; i.e. an \emph{ordered tuple} of indices.\\    
$|\vec{\alpha}|$ & $\equiv$ & $k$, the number of elements in the tuple $\vec{\alpha} = \left(\alpha_1, \alpha_2, \dots, \alpha_k \right)$.\\      
$[k]$ & $\equiv$ & $ (1, 2, \dots, k)$, the particular tuple consisting of consecutive integers from $1$ to $k$.\\
$\overline{\alpha}$ & $\equiv$ & Ordered tuple of indices in the set-complement of $\vec{\alpha}$ in $[N]$. \\ 
$()$ & $\equiv$ & The empty tuple, for which $|()| = 0$. \\
$\vec{\alpha} \cup \vec{\beta}$ & $\equiv$ & Ordered tuple of all indices in either $\vec{\alpha}$ or $\vec{\beta}$. \\
$\vec{\alpha} \cap \vec{\beta}$ & $\equiv$ & Ordered tuple of all indices in both $\vec{\alpha}$ and $\vec{\beta}$. \\
$\vec{\alpha} \subset \vec{\beta}$ & $\equiv$ & $\vec{\alpha}$ is an ordered tuple of indices $\alpha_j \in \vec{\beta}$ for all $j \in \{1, 2, \dots, |\vec{\alpha}| < |\vec{\beta}|\}$. \\
$\vec{\alpha}\backslash \alpha_j$ & $\equiv$ & Ordered tuple of indices from $\vec{\alpha}$ with the particular element $\alpha_j$ removed. \\
$(\vec{\alpha}, \vec{\beta})$ & $\equiv$ & $(\alpha_1, \alpha_2, \dots, \alpha_k, \beta_1, \beta_2, \dots)$; i.e. the unordered concatenation of $\vec{\alpha}$ and $\vec{\beta}$. \\
$C_{\vec{\alpha}}$ & $\equiv$ & $ \prod_{j = 1}^{|\vec{\alpha}|} c_{\alpha_j}$, the product of Majorana operators with indices in ascending order, called a \emph{Majorana configuration operator}. \\
$\mathbf{u}_{\vec{\alpha} \vec{\beta}}$ & $\equiv$ & Submatrix $\mathbf{u}_{\vec{\alpha} \vec{\beta}}$ of $\mathbf{u}$, with row indices from $\vec{\alpha}$ and column indices from $\vec{\beta}$; i.e. $\left(\mathbf{u}_{\vec{\alpha} \vec{\beta}}\right)_{jk} = u_{\alpha_j \beta_k}$. 
\end{tabular}
\end{center}
\label{tab:notation}
\end{table}

\noindent Additionally, when we say we have a matchgate unitary $U$ on $n$ qubits, we associate it with the $2n \times 2n$ orthogonal matrix $\mathbf{u} \in \mathbf{SO}(2n)$ by the corresponding lowercase symbol in bold. We also denote block matrices in the usual way [e.g. $\left(\begin{array}{c c} 
\mathbf{A} & \mathbf{B} \\ \mathbf{C} & \mathbf{D}
\end{array}\right)$].

\section{Appendix A: Majorana Configuration Operator Transition Amplitudes}
\label{sec:CCAmps}

Here we prove the formula

\begin{equation}
U^{\dagger} C_{\vec{\alpha}} U = \sum_{\{\vec{\beta} | |\vec{\beta}| = |\vec{\alpha}| \}} \det \left(\mathbf{u}_{\vec{\alpha} \vec{\beta}} \right) C_{\vec{\beta}} \rm{,}
\label{eq:MMTransition}
\end{equation}

\noindent for some matchgate unitary $U$, by induction on the number of Majorana factors $k \equiv |\vec{\alpha}|$. First consider the case where $k = 2$, and let $\alpha_1 < \alpha_2$. We have

\begin{align*}
U^{\dagger} c_{\alpha_1} c_{\alpha_2} U &= \left(U^{\dagger} c_{\alpha_1} U \right) \left(U^{\dagger} c_{\alpha_2} U \right) \\
&= \left(\sum_{\beta_1} u_{\alpha_1 \beta_1} c_{\beta_1} \right) \left(\sum_{\beta_2} u_{\alpha_2 \beta_2} c_{\beta_2} \right) \\
&= \left(\sum_{\{(\beta_1, \beta_2) | \beta_1 = \beta_2\}} + \sum_{\{(\beta_1, \beta_2) | \beta_1 < \beta_2\}} + \sum_{\{(\beta_1, \beta_2) | \beta_1 > \beta_2\}} \right) u_{\alpha_1 \beta_1} u_{\alpha_2 \beta_2} c_{\beta_1} c_{\beta_2} \\
&= \left(\sum_{\beta_1} u_{\alpha_1 \beta_1} u_{\alpha_2 \beta_1} \right) I +  \sum_{\{(\beta_1, \beta_2) | \beta_1 < \beta_2\}} \left( u_{\alpha_1 \beta_1} u_{\alpha_2 \beta_2} -  u_{\alpha_1 \beta_2} u_{\alpha_2 \beta_1} \right) c_{\beta_1} c_{\beta_2} \\
&= \sum_{\{(\beta_1, \beta_2) | \beta_1 < \beta_2\}} \det\left[ \mathbf{u}_{\left(\alpha_1, \alpha_2 \right) \left(\beta_1, \beta_2 \right)} \right] c_{\beta_1} c_{\beta_2} \\ \rm{,}
\end{align*}

\noindent where, from the third to the fourth line, we used the Majorana algebra anticommutation relations

\begin{align*}
\{c_{\mu}, c_{\nu} \} = 2\delta_{\mu \nu} I \rm{,}
\end{align*}

\noindent relabeling dummy indices $\beta_1 \leftrightarrow \beta_2$ on the third sum in the third line. From the fourth to the fifth line, we see that the identity term vanishes as its coefficient is the inner product between two distinct row vectors of an orthogonal matrix. This proves the statement for $k = 2$. Next, we assume the statement holds for general $k$ and use this assumption to prove the statement for $k + 1$. Without loss of generality, assume $\alpha_j < \alpha_{k + 1}$ for all $j \leq k$. We now have

\setcounter{Hequation}{1}
\renewcommand{\theHequation}{S\arabic{Hequation}}

\begin{align}
U^{\dagger} C_{\vec{\alpha}} c_{\alpha_{k + 1}} U &= \left( U^{\dagger} C_{\vec{\alpha}} U \right) \left( U^{\dagger} c_{\alpha_{k + 1}} U  \right) \nonumber \\
&= \left[\sum_{\{\vec{\beta}| |\vec{\beta}| = |\vec{\alpha}|\}} \det\left( \mathbf{u}_{\vec{\alpha} \vec{\beta} } \right) C_{\vec{\beta}} \right] \left(\sum_{\beta_{k + 1}} u_{\alpha_{k +1} \beta_{k + 1}} c_{\beta_{k + 1}} \right) \nonumber \\
U^{\dagger} C_{\vec{\alpha}} c_{\alpha_{k + 1}} U &= \sum_{\{(\vec{\beta}, \beta_{k + 1}) | |\vec{\beta}| = |\vec{\alpha}|\}} u_{\alpha_{k + 1} \beta_{k + 1}} \det\left( \mathbf{u}_{\vec{\alpha} \vec{\beta}} \right) C_{\vec{\beta}}  c_{\beta_{k + 1}} \label{eq:indsum}
\end{align}

\noindent  Each of the terms in the sum above falls into one of two categories. Either (i) $\beta_{k + 1} \in \vec{\beta}$, and $C_{\vec{\beta}} c_{\beta_{k + 1}} = \pm C_{\vec{\beta}\backslash\beta_{k + 1}}$, or (ii) $\beta_{k + 1} \notin \vec{\beta}$, and $C_{\vec{\beta}} c_{\beta_{k + 1}} = \pm C_{\vec{\beta} \cup \beta_{k + 1}}$, with sign given  in both cases by  $(-1)^{|\{j \leq k | \beta_{j} > \beta_{k + 1}\}|}$. We first proceed to demonstrate that all of the terms in category (i) vanish. Fix a particular configuration operator $C_{\vec{\beta} \backslash \beta_{k + 1}}$. The coefficient on this operator in the r.h.s. of Eq.~(\ref{eq:indsum}) is given by a sum over all indices $\gamma$ which could have been removed from $\vec{\beta}$ to yield $\vec{\beta} \backslash \beta_{k + 1}$

\begin{align*}
\sum_{\gamma \notin \vec{\beta}\backslash\beta_{k + 1}} u_{\alpha_{k + 1} \gamma} \det\left[\mathbf{u}_{\vec{\alpha} \left( \vec{\beta}\backslash\beta_{k + 1}, \gamma \right)} \right] = \sum_{\gamma} u_{\alpha_{k + 1} \gamma} \det\left[\mathbf{u}_{\vec{\alpha} \left( \vec{\beta}\backslash\beta_{k + 1}, \gamma \right)} \right] \rm{,}
\end{align*}

\noindent where we were able to cancel any sign factors on the terms inside the sum by reordering columns in $\mathbf{u}$ so that the $\gamma$ column appears at the rightmost position, using the alternating sign property of the determinant. The equality is due to the fact that if $\gamma \in \vec{\beta}\backslash\beta_{k + 1}$, then the determinant in that term evaluates to zero. Finally, we use the multilinearity property of the determinant to bring the sum on to the last column, as  

\setcounter{Hequation}{2}
\renewcommand{\theHequation}{S\arabic{Hequation}}

\begin{align}
\sum_{\gamma} u_{\alpha_{k + 1} \gamma} \det\left[\mathbf{u}_{\vec{\alpha} \left( \vec{\beta}\backslash\beta_{k + 1}, \gamma \right)} \right] &= \det \left[\left(
\begin{array}{c c} 
\mathbf{u}_{\vec{\alpha}, \vec{\beta}\backslash\beta_{k + 1}} & \sum_{\gamma} u_{\alpha_{k + 1} \gamma} \mathbf{u}_{\vec{\alpha} \gamma} 
\end{array} 
\right) \right] \mathrm{,}
\label{eq:multilinear}
\end{align}

\noindent i.e. the determinant of a matrix whose last column vector is $\sum_{\gamma} u_{\alpha_{k + 1} \gamma} \mathbf{u}_{\vec{\alpha} \gamma}$. The $l$th element of this column is given by

\begin{align*}
\sum_{\gamma} u_{\alpha_{k + 1} \gamma} u_{\alpha_l \gamma} = \delta_{\alpha_{k + 1} \alpha_{l}}
\end{align*}

\noindent again following from the fact that this sum is the inner product between two column vectors of an orthogonal matrix. However, $\alpha_{k + 1} > \alpha_l$ by assumption, so this sum is actually always zero and the determinant in (\ref{eq:multilinear}) vanishes. Each of the terms in category (i) therefore vanishes, and the only terms in the r.h.s of Eq.~(\ref{eq:indsum}) that survive are in category (ii). We examine these terms by next fixing a particular configuration operator $C_{\vec{\beta} \cup \beta_{k + 1}}$. The coefficient on this operator in the r.h.s. of (\ref{eq:indsum}) is given by a sum over all indices $\gamma$ that could have been added to $\vec{\beta}$ to yield $\vec{\beta} \cup \beta_{k + 1}$ (we cannot cancel sign factors this time)

\begin{align*}
\sum_{\gamma \in \vec{\beta} \cup \beta_{k + 1}} (-1)^{|\{j \leq k + 1 | \beta_j > \gamma \}|} u_{\alpha_{k + 1} \gamma} \det \left[\mathbf{u}_{\vec{\alpha},\left(\vec{\beta} \cup \beta_{k + 1}\right)\backslash\gamma} \right] &= \det \left[\mathbf{u}_{(\vec{\alpha}, \alpha_{k + 1}), \vec{\beta} \cup \beta_{k + 1}} \right] \rm{.}
\end{align*}

\noindent  To see that this equality indeed holds, relabel indices in $\vec{\beta} \cup \beta_{k + 1}$ such that $\beta_i < \beta_{i + 1}$ for all $i \leq k$, and suppose $\gamma = \beta_s$ in this labeling. Then we have

\begin{align*}
(-1)^{|\{j \leq k + 1 | \beta_j > \gamma \}|} = (-1)^{(k + 1) - s} = (-1)^{(k + 1) + s} \\
\end{align*} 

\noindent As $\alpha_j < \alpha_{k + 1}$ for all $j \leq k$, this is exactly the sign factor that would appear had we expanded along the last [$(k + 1)$st] row of the matrix in the r.h.s. above, since $\beta_s$ appears as the $s$th column of this matrix. We therefore have

\begin{align*}
U^{\dagger} C_{\vec{\alpha}} c_{\alpha_{k + 1}} U &=  \sum_{\{\vec{\beta} | |\vec{\beta}| = k + 1\}} \det \left[\mathbf{u}_{(\vec{\alpha}, \alpha_{k + 1}) \vec{\beta}} \right] C_{\vec{\beta}} \rm{,}
\end{align*}

\noindent which proves the statement for $k + 1$, given that it holds for $k$. This completes our inductive proof of Eq.~(\ref{eq:MMTransition}). $\square$

\section{Appendix B: Proof of Theorem 1}
\label{sec:ProofThm1}

Let $A$ and $B$ be two Pauli operators (i.e. $A^{\dagger} = A$, $B^{\dagger} = B$, and $A^2= B^2 = I$) and $U$ be a matchgate unitary, then we have

\begin{align*}
||[A, U^{\dagger} B U]||^2 &= \tr\{[A, U^{\dagger} B U]^{\dagger}[A, U^{\dagger} B U]\} \\
&= -\tr\{[A, U^{\dagger} B U]^2\} \\
&= -\tr\left[A \left(U^{\dagger} B U\right) A \left(U^{\dagger} B U\right) - A \left(U^{\dagger} B U\right) \left(U^{\dagger} B U\right) A  - \left(U^{\dagger} B U\right) A A \left(U^{\dagger} B U\right) + \left(U^{\dagger} B U\right) A \left(U^{\dagger} B U\right) A \right] \\
&= \tr\left[2 I - 2 A \left(U^{\dagger} B U\right) A \left(U^{\dagger} B U\right) \right] \\
||[A, U^{\dagger} B U]||^2 &= 2 \left\{2^n - \tr\left[A \left(U^{\dagger} B U\right) A \left(U^{\dagger} B U\right)\right]\right\}
\end{align*}

\noindent From the first to the second line, we used Hermiticity of $A$ and $B$, and we expanded the expresion in the third line. From the third to the fourth line, we used $A^2= B^2 = I$, and from the fourth to the fifth line, we used $\tr(I) = 2^n$ on $n$ qubits. Dividing by a normalization of $2^{n + 2}$ on both sides, we have

\setcounter{Hequation}{3}
\renewcommand{\theHequation}{S\arabic{Hequation}}

\begin{align}
\frac{1}{2^{n + 2}}||[A, U^{\dagger} B U]||^2 = \frac{1}{2} \left\{1 - \frac{1}{2^n}\tr\left[A \left(U^{\dagger} B U\right) A \left(U^{\dagger} B U\right)\right]\right\}
\label{eq:paulithm1}
\end{align}

\noindent Next, we assume $A = i^{a} C_{\vec{\eta}}$ and $B = i^{b} C_{\vec{\alpha}}$ for some integers $a$ and $b$, such that $i^a$ and $i^b$ ensure that $A$ and $B$ are Hermitian, respectively. E.g., for $A$, this implies

\setcounter{Hequation}{4}
\renewcommand{\theHequation}{S\arabic{Hequation}}

\begin{align}
A^{\dagger} &= (-i)^a C^{\dagger}_{\vec{\eta}} \nonumber \\
A^{\dagger} &= (-i)^a (-1)^{\frac{1}{2}|\vec{\eta}| (|\vec{\eta}| - 1)} C_{\vec{\eta}} \label{eq:reverse} \rm{,}
\end{align}

\noindent where $C^{\dagger}_{\vec{\eta}} = (-1)^{\frac{1}{2}|\vec{\eta}| (|\vec{\eta}| - 1)} C_{\vec{\eta}}$, since the individual Majorana modes are Hermitian and we need $\frac{1}{2}|\vec{\eta}| (|\vec{\eta}| - 1)$ anticommutations between individual distinct modes to reverse the order of $|\vec{\eta}|$ modes. Since $A^{\dagger} = A$, Eq.~(\ref{eq:reverse}) implies

\begin{align*}
i^{a} C_{\vec{\eta}} &= (-i)^a (-1)^{\frac{1}{2}|\vec{\eta}| (|\vec{\eta}| - 1)} C_{\vec{\eta}} \rm{,}
\end{align*}	

\noindent and multiplying both sides of this equation by $i^a C^{\dagger}_{\vec{\eta}}$ gives 

\begin{align*}
i^{2a} I &= (-1)^{\frac{1}{2}|\vec{\eta}| (|\vec{\eta}| - 1)} I \rm{.}
\end{align*}	

\noindent Thus, $i^{2a} = (-1)^{\frac{1}{2}|\vec{\eta}| (|\vec{\eta}| - 1)}$, and similarly, $i^{2b} = (-1)^{\frac{1}{2}|\vec{\alpha}| (|\vec{\alpha}| - 1)}$. Continuing from the dynamical term in Eq.~(\ref{eq:paulithm1}), we have
	
\begin{align*}
& \frac{1}{2^n} (-1)^{\frac{1}{2} \left[|\vec{\alpha}|(|\vec{\alpha}| - 1) + |\vec{\eta}|(|\vec{\eta}| - 1)  \right]} \tr\left[A \left(U^\dagger B U\right) A \left(U^\dagger B U\right) \right] = \frac{1}{2^n} \tr\left[C_{\vec{\eta}} \left(U^\dagger C_{\vec{\alpha}} U\right) C_{\vec{\eta}} \left(U^\dagger C_{\vec{\alpha}} U\right) \right] \\
&\mbox{\hspace{20mm}}= \frac{1}{2^n} \sum_{\{\vec{\beta}, \vec{\beta}'  | |\vec{\beta}| = |\vec{\beta}'| = |\vec{\alpha}|\}} \det \left(\mathbf{u}_{\vec{\alpha} \vec{\beta}}\right) \det \left(\mathbf{u}_{\vec{\alpha} \vec{\beta}'}\right) \tr \left(C_{\vec{\eta}} C_{\vec{\beta}} C_{\vec{\eta}} C_{\vec{\beta}'} \right) \\
& \mbox{\hspace{20mm}}= \frac{1}{2^n} (-1)^{|\vec{\alpha}||\vec{\eta}| + \frac{1}{2} |\vec{\eta}|(|\vec{\eta}| - 1)} \sum_{\{\vec{\beta}, \vec{\beta}'  | |\vec{\beta}| = |\vec{\beta}'| = |\vec{\alpha}|\}} (-1)^{|\vec{\eta} \cap \vec{\beta}|} \det \left(\mathbf{u}_{\vec{\alpha} \vec{\beta}}\right) \det \left(\mathbf{u}_{\vec{\alpha} \vec{\beta}'}\right) \tr \left(C_{\vec{\beta}} C_{\vec{\beta}'} \right) \\
& \mbox{\hspace{20mm}}= (-1)^{|\vec{\alpha}||\vec{\eta}| + \frac{1}{2} \left[|\vec{\alpha}|(|\vec{\alpha}| - 1) + |\vec{\eta}|(|\vec{\eta}| - 1)  \right]} \sum_{\{\vec{\beta}, \vec{\beta}'  | |\vec{\beta}| = |\vec{\beta}'| = |\vec{\alpha}|\}} (-1)^{|\vec{\eta} \cap \vec{\beta}|} \det \left(\mathbf{u}_{\vec{\alpha} \vec{\beta}}\right) \det \left(\mathbf{u}_{\vec{\alpha} \vec{\beta}'}\right) \delta_{\vec{\beta}\vec{\beta}'} \\
& \frac{1}{2^n}\tr\left[A \left(U^\dagger B U\right) A \left(U^\dagger B U\right) \right] = (-1)^{|\vec{\alpha}||\vec{\eta}|} \sum_{\{\vec{\beta}, \vec{\beta}'  | |\vec{\beta}| = |\vec{\beta}'| = |\vec{\alpha}|\}} (-1)^{|\vec{\eta} \cap \vec{\beta}|} \det \left(\mathbf{u}_{\vec{\alpha} \vec{\beta}}\right) \det \left(\mathbf{u}_{\vec{\alpha} \vec{\beta}'}\right) \delta_{\vec{\beta}\vec{\beta}'}
\end{align*} 

\noindent From the second to the third line, we used 

\begin{align*}
C_{\vec{\eta}}C_{\vec{\beta}} &= (-1)^{|\vec{\beta}||\vec{\eta}| + |\vec{\eta} \cap \vec{\beta}|} C_{\vec{\beta}} C_{\vec{\eta}} \\
C_{\vec{\eta}}^2 &= (-1)^{\frac{1}{2} |\vec{\eta}| (|\vec{\eta}| - 1)} I
\end{align*}

\noindent and similarly from the third to the fourth line, as well as the fact that $|\vec{\beta}| = |\vec{\alpha}|$. We recognize that 

\begin{align*}
\det\left[(\mathbf{I} - 2 \mathbf{P}_{\vec{\eta}})_{\vec{\beta} \vec{\beta}'}\right] = (-1)^{|\vec{\eta} \cap \vec{\beta}|} \delta_{\vec{\beta} \vec{\beta}'}
\end{align*}

\noindent where $\mathbf{P}_{\vec{\eta}}$ is the projector onto modes $\vec{\eta}$. We therefore have

\begin{align*}
\frac{1}{2^n}\tr\left[A \left(U^\dagger B U\right) A \left(U^\dagger B U\right) \right] &= (-1)^{|\vec{\alpha}||\vec{\eta}|} \sum_{\{\vec{\beta}, \vec{\beta}'  | |\vec{\beta}| = |\vec{\beta}'| = |\vec{\alpha}|\}} \det \left(\mathbf{u}_{\vec{\alpha} \vec{\beta}}\right) \det\left[(\mathbf{I} - 2 \mathbf{P}_{\vec{\eta}})_{\vec{\beta} \vec{\beta}'}\right] \det \left(\mathbf{u}_{\vec{\alpha} \vec{\beta}'}\right) \\
\frac{1}{2^n}\tr\left[A \left(U^\dagger B U\right) A \left(U^\dagger B U\right) \right] &= (-1)^{|\vec{\alpha}||\vec{\eta}|} \det [\mathbf{u}_{\vec{\alpha} [2n]} (\mathbf{I} - 2\mathbf{P}_{\vec{\eta}}) \mathbf{u}^{\mathrm{T}}_{[2n] \vec{\alpha}}] \rm{,}
\end{align*}

\noindent which follows from the Cauchy-Binet formula. This therefore proves the theorem

\begin{align}
\frac{1}{2^{n + 2}}||[A, U^{\dagger} B U]||^2 = \frac{1}{2} \left\{1 - (-1)^{|\vec{\alpha}||\vec{\eta}|} \det [\mathbf{u}_{\vec{\alpha} [2n]} (\mathbf{I} - 2\mathbf{P}_{\vec{\eta}}) \mathbf{u}^{\mathrm{T}}_{[2n] \vec{\alpha}}] \right\}
\label{eq:majthm1}
\end{align}

\noindent $\square$

\setcounter{Hequation}{5}
\renewcommand{\theHequation}{S\arabic{Hequation}}

\section{Appendix C: Modified Cauchy-Binet Formula}
\label{sec:MCBForm}

Here we prove 

\setcounter{Hequation}{6}
\renewcommand{\theHequation}{S\arabic{Hequation}}

\begin{align}
\sum_{\{\vec{\beta} \subset \vec{B} | |\vec{\beta}| = |\vec{\alpha}| - |\vec{S}|\}} \det \left(\mathbf{u}_{\vec{\alpha}, \vec{\beta} \cup \vec{S}} \right)
\det \left(\mathbf{v}_{\vec{\alpha}, \vec{\beta} \cup \vec{S}} \right) = (-1)^{|\vec{S}|} \det
\left(
\begin{array}{c c}
  \mathbf{0}_{|\vec{S}| \times |\vec{S}|} & \mathbf{v}^{\rm{T}}_{\vec{S} \vec{\alpha}} \\
  \mathbf{u}_{\vec{\alpha} \vec{S}} & \mathbf{u}_{\vec{\alpha} \vec{B}} \mathbf{v}^{\rm{T}}_{\vec{B} \vec{\alpha}}  \end{array}
\right)
\label{eq:modifiedCB}
\end{align}

\noindent for $\vec{S}$ disjoint from $\vec{B}$. We first rearrange rows and columns inside the matrices $\mathbf{u}$ and $\mathbf{v}$ in the l.h.s. of (\ref{eq:modifiedCB}) to bring each of them to a fiducial form, $\mathbf{u}^{\prime}$ and $\mathbf{v}^{\prime}$, respectively. These are such that $\mathbf{u}^{\prime}_{\vec{\alpha}^{\prime} \vec{S}^{\prime}} = \mathbf{u}_{\vec{\alpha}\vec{S}}$ and $\mathbf{u}^{\prime}_{\vec{\alpha}^{\prime} \vec{B}^{\prime}} = \mathbf{u}_{\vec{\alpha}\vec{B}}$ (and similarly for $\mathbf{v}^{\prime}$), for $\vec{\alpha}^{\prime} \equiv [|\vec{\alpha}|]$ and $\vec{S}^{\prime} \equiv [|\vec{S}|]$. That is, we bring the rows $\vec{\alpha}$ to the top and the columns $\vec{S}$ to the left inside the matrices $\mathbf{u}$ and $\mathbf{v}$ without changing the internal ordering of these tuples, nor the ordering of $\vec{B}$. This is done purely for convenience of presentation and will not affect the argument, as we will undo the rearrangement in the end. We will continue to refer to the numbers of elements in these rearranged tuples by those of their unprimed counterparts (i.e. using $|\vec{S}|$ instead of $|\vec{S}^{\prime}|$), as they are equal. Since this rearrangement is done for both $\mathbf{u}$ and $\mathbf{v}$, any resulting sign factor acquired due to the alternating sign property of the determinant will cancel in the product, and we have

\setcounter{Hequation}{7}
\renewcommand{\theHequation}{S\arabic{Hequation}}

\begin{align}
\sum_{\{\vec{\beta} \subset \vec{B} | |\vec{\beta}| = |\vec{\alpha}| - |\vec{S}|\}} \det \left(\mathbf{u}_{\vec{\alpha}, \vec{\beta} \cup \vec{S}} \right)
\det \left(\mathbf{v}_{\vec{\alpha}, \vec{\beta} \cup \vec{S}} \right) &= \sum_{\{\vec{\beta} \subset \vec{B}^{\prime} | |\vec{\beta}| = |\vec{\alpha}| - |\vec{S}|\}} \det \left[\mathbf{u}^{\prime}_{\vec{\alpha}^{\prime} (\vec{S}^{\prime}, \vec{\beta})} \right]
\det \left[\mathbf{v}^{\prime}_{\vec{\alpha}^{\prime} (\vec{S}^{\prime}, \vec{\beta})} \right] \rm{,}
\label{eq:modifiedCB2}
\end{align}

\setcounter{Hequation}{8}
\renewcommand{\theHequation}{S\arabic{Hequation}}

\noindent We will next need the Laplace ``expansion by complimentary minors" formula

\begin{align}
\det \left(\mathbf{u}\right) = \sum_{\{\vec{H} | |\vec{H}| = k\}} \varepsilon^{\vec{H}, \vec{L}} \det \left( \mathbf{u}_{\vec{H} \vec{L}} \right) \det \left( \mathbf{u}_{\bar{H} \bar{L}} \right) \rm{,}
\label{eq:Laplacecomp}
\end{align}

\noindent where $\varepsilon^{\vec{H}, \vec{L}} = (-1)^{\sum_{j = 1}^k \left(H_{j} + L_{j}\right)}$, $\vec{L}$ is a fixed subset of the columns of $\mathbf{u}$ of size $k$, the sum is over all subsets $\vec{H}$ of rows of $\mathbf{u}$ of size $k$, and $\overline{L}$ and $\overline{H}$ are the set-complements of $\vec{L}$ and $\vec{H}$ in the sets of all columns and rows of $\mathbf{u}$, respectively. This is the analogous formula to expanding the determinant by minors of a fixed column, generalized to a subset of columns $\vec{L}$. Applying Eq.~(\ref{eq:Laplacecomp}) to the columns $\vec{S}^{\prime}$ of $\mathbf{u}^{\prime}$ and $\mathbf{v}^{\prime}$ on the r.h.s. of (\ref{eq:modifiedCB2}) gives

\begin{align*}
\sum_{\{\vec{\beta} \subset \vec{B} | |\vec{\beta}| = |\vec{\alpha}| - |\vec{S}|\}} \det \left(\mathbf{u}_{\vec{\alpha}, \vec{\beta} \cup \vec{S}} \right)
\det \left(\mathbf{v}_{\vec{\alpha}, \vec{\beta} \cup \vec{S}} \right) &=\sum_{\{\vec{\beta} \subset \vec{B}^{\prime} | |\vec{\beta}| = |\vec{\alpha}| - |\vec{S}|\}} \left[ \sum_{\{\vec{H} \subset \vec{\alpha}^{\prime} | |\vec{H}| = |\vec{S}|\}}  \varepsilon^{\vec{H}, \vec{S}^{\prime}} \det \left( \mathbf{u}^\prime_{\vec{H} \vec{S}^{\prime}} \right) \det \left( \mathbf{u}^{\prime}_{\vec{\alpha}^{\prime}\backslash\vec{H}, \vec{\beta}} \right) \right] \\
& \hspace{15mm} \times \left[\sum_{\{\vec{L} \subset \vec{\alpha}^{\prime} | |\vec{L}| = |\vec{S}|\}} \varepsilon^{\vec{L}, \vec{S}^{\prime}} \det \left( \mathbf{v}^{\prime}_{\vec{L} \vec{S}^{\prime}} \right) \det \left( \mathbf{v}^{\prime}_{ \vec{\alpha}^{\prime}\backslash\vec{L}, \vec{\beta}} \right) \right] \\
&= \sum_{\substack{\{\vec{H} \subset \vec{\alpha}^{\prime}| |\vec{H}| = |\vec{S}| \} \\ \{\vec{L} \subset\vec{\alpha}^{\prime}| |\vec{L}| = |\vec{S}| \}}}  \varepsilon^{\vec{H}, \vec{S}^{\prime}}  \varepsilon^{\vec{L}, \vec{S}^{\prime}} \det \left( \mathbf{u}^{\prime}_{\vec{H} \vec{S}^{\prime}} \right) \det \left( \mathbf{v}^{\prime}_{\vec{L} \vec{S}^{\prime}} \right) \\ 
& \hspace{15mm} \times \left[ \sum_{\{\vec{\beta} \subset \vec{B}^{\prime} | |\vec{\beta}| = |\vec{\alpha}| - |\vec{S}| \}} \det \left(\mathbf{u}^{\prime}_{\vec{\alpha}^{\prime}\backslash\vec{H}, \vec{\beta}} \right) \det \left( \mathbf{v}^{\prime}_{\vec{\alpha}^{\prime}\backslash\vec{L}, \vec{\beta}} \right) \right] \rm{.}
\end{align*}

\noindent We next apply the Cauchy-Binet formula to the sum in square brackets, as

\setcounter{Hequation}{9}
\renewcommand{\theHequation}{S\arabic{Hequation}}

\begin{align}
 \sum_{\{\vec{\beta} \subset \vec{B}^{\prime} | |\vec{\beta}| = |\vec{\alpha}| - |\vec{S}| \}} \det \left(\mathbf{u}^{\prime}_{\vec{\alpha}^{\prime}\backslash\vec{H}, \vec{\beta}} \right) \det \left( \mathbf{v}^{\prime}_{\vec{\alpha}^{\prime}\backslash\vec{L}, \vec{\beta}} \right) = \det \left( \mathbf{u}^{\prime}_{\vec{\alpha}^{\prime}\backslash\vec{H}, \vec{B}^{\prime} } \mathbf{v}^{\prime \rm{T}}_{\vec{B}^{\prime}, \vec{\alpha}^{\prime}\backslash\vec{L}} \right) \rm{.}
\label{eq:partialCB}
\end{align}

\noindent Notice that the matrix in the determinant of the r.h.s. above is simply $\mathbf{u}^{\prime}_{\vec{\alpha}^{\prime} \vec{B}^{\prime} } \mathbf{v}^{\prime \rm{T}}_{\vec{B}^{\prime} \vec{\alpha}^{\prime}}$ with the rows $\vec{H}$ and columns $\vec{L}$ removed (i.e. instead of removing the rows and columns and then multiplying, we can multiply and then remove rows and columns from the product). This gives

\setcounter{Hequation}{10}
\renewcommand{\theHequation}{S\arabic{Hequation}}

\begin{align}
\sum_{\{\vec{\beta} \subset \vec{B} | |\vec{\beta}| = |\vec{\alpha}| - |\vec{S}|\}} \det \left(\mathbf{u}_{\vec{\alpha}, \vec{\beta} \cup \vec{S}} \right)
\det \left(\mathbf{v}_{\vec{\alpha}, \vec{\beta} \cup \vec{S}} \right) &= \sum_{\substack{\{\vec{H} \subset \vec{\alpha}^{\prime}| |\vec{H}| = |\vec{S}| \} \\ \{\vec{L} \subset \vec{\alpha}^{\prime}| |\vec{L}| = |\vec{S}| \}}}  \varepsilon^{\vec{H}, \vec{S}^{\prime}}  \varepsilon^{\vec{L}, \vec{S}^{\prime}} \det \left( \mathbf{u}^{\prime}_{\vec{H} \vec{S}^{\prime}} \right) \det \left( \mathbf{v}^{\prime}_{\vec{L} \vec{S}^{\prime}} \right) \nonumber \\
& \hspace{15mm} \times \det \left[ \left( \mathbf{u}^{\prime}_{\vec{\alpha}^{\prime} \vec{B}^{\prime} } \mathbf{v}^{\prime \rm{T}}_{\vec{B}^{\prime} \vec{\alpha}^{\prime}} \right)_{\vec{\alpha}^{\prime}\backslash\vec{H}, \vec{\alpha}^{\prime}\backslash\vec{L}} \right] \rm{.}
\label{eq:modifiedCB3}
\end{align}
 
\noindent The next step is to ``put back in" the rows $\vec{H}$ and columns $\vec{L}$. We do this by treating (\ref{eq:modifiedCB3}) as an expansion of the determinant of a larger matrix by complimentary minors of rows $\mathbf{v}^{\prime \rm{T}}_{\vec{S}^{\prime} \vec{\alpha}^{\prime}}$ and then columns $\mathbf{u}^{\prime}_{\vec{\alpha}^{\prime} \vec{S}^{\prime}}$. Working backwards, we see that the sum over $\vec{H}$ in Eq.~(\ref{eq:modifiedCB3}) evaluates to 

\begin{align*}
\sum_{\{\vec{H} \subset \vec{\alpha}^{\prime} | |\vec{H}| = |\vec{S}|\}} \varepsilon^{\vec{H}, \vec{S}^{\prime}} \det \left(\mathbf{u}^{\prime}_{\vec{H} \vec{S}^{\prime}} \right) \det \left[ \left( \mathbf{u}^{\prime}_{\vec{\alpha}^{\prime} \vec{B}^{\prime} } \mathbf{v}^{\prime \rm{T}}_{\vec{B}^{\prime} \vec{\alpha}^{\prime}} \right)_{\vec{\alpha}^{\prime}\backslash\vec{H}, \vec{\alpha}^{\prime}\backslash\vec{L}} \right] \\ 
\hspace{15mm} = \det \left[
\left(
\begin{array}{c c}
  \mathbf{u}^{\prime}_{\vec{\alpha}^{\prime} \vec{S}^{\prime}} & \left( \mathbf{u}^{\prime}_{\vec{\alpha}^{\prime} \vec{B}^{\prime} } \mathbf{v}^{\prime \rm{T}}_{\vec{B}^{\prime} \vec{\alpha}^{\prime}} \right)_{\vec{\alpha}^{\prime}, \vec{\alpha}^{\prime}\backslash\vec{L}}  \end{array}
\right) \right]
\end{align*}

\noindent To put the rows back in, we note that

\setcounter{Hequation}{11}
\renewcommand{\theHequation}{S\arabic{Hequation}}

\begin{align}
\det \left(\mathbf{v}^{\prime}_{\vec{L} \vec{S}^{\prime}}\right) = \det \left(\mathbf{v}^{\prime \rm{T}}_{\vec{S}^{\prime} \vec{L}}\right) = (-1)^{|\vec{S}|} \det \left[
\left(
\begin{array}{c c}
  \mathbf{0}_{|\vec{S}| \times |\vec{S}|} & \mathbf{v}^{\prime \rm{T}}_{\vec{S}^{\prime} \vec{\alpha}^{\prime}}  \end{array}
\right)_{\vec{S}^{\prime} \vec{L}^{\prime}} \right] \rm{,}
\label{eq:modifiedCB4}
\end{align}

\noindent where $\vec{L}^{\prime}$ is related to $\vec{L}$ by $L^{\prime}_j = L_j + |\vec{S}|$ for all $j \in \{1, 2, \dots, |\vec{L}| = |\vec{S}|\}$. Shifting the columns over by $|\vec{S}|$ inside the determinant gives the overall factor of $(-1)^{|\vec{S}|}$. Thus, the sum over $\vec{L}$ in Eq.~(\ref{eq:modifiedCB3}) evaluates to

\begin{align*}
&\sum_{\{\vec{L} \subset \vec{\alpha}^{\prime}| |\vec{L}| = |\vec{S}| \}}  \varepsilon^{\vec{L}, \vec{S}^{\prime}} \det \left( \mathbf{v}^{\prime}_{\vec{L} \vec{S}^{\prime}} \right)\det \left[
\left(
\begin{array}{c c}
  \mathbf{u}^{\prime}_{\vec{\alpha}^{\prime} \vec{S}^{\prime}} & \left( \mathbf{u}^{\prime}_{\vec{\alpha}^{\prime} \vec{B}^{\prime} } \mathbf{v}^{\prime \rm{T}}_{\vec{B}^{\prime} \vec{\alpha}^{\prime}} \right)_{\vec{\alpha}^{\prime}, \vec{\alpha}^{\prime}\backslash\vec{L}}  \end{array}
\right) \right] \\ 
&\hspace{15mm}= (-1)^{|\vec{S}|} \sum_{\{\vec{L} \subset \left[|\vec{\alpha}| + |\vec{S}| \right]| |\vec{L}| = |\vec{S}| \}}  \varepsilon^{\vec{L}, \vec{S}^{\prime}} \det \left[
\left(
\begin{array}{c c}
  \mathbf{0}_{|\vec{S}| \times |\vec{S}|} & \mathbf{v}^{\prime \rm{T}}_{\vec{S}^{\prime} \vec{\alpha}^{\prime}}  \end{array}
\right)_{\vec{S}^{\prime} \vec{L}} \right] \\ 
&\hspace{30mm}\times \det \left[
\left(
\begin{array}{c c}
  \mathbf{u}^{\prime}_{\vec{\alpha}^{\prime} \vec{S}^{\prime}} & \mathbf{u}^{\prime}_{\vec{\alpha}^{\prime} \vec{B}^{\prime} } \mathbf{v}^{\prime \rm{T}}_{\vec{B}^{\prime} \vec{\alpha}^{\prime}}  \end{array}
\right)_{\vec{\alpha}^{\prime}, \vec{\alpha}^{\prime}\backslash\vec{L}} \right] \\
&\hspace{15mm}= (-1)^{|\vec{S}|} \det
\left(
\begin{array}{c c}
  \mathbf{0}_{|\vec{S}| \times |\vec{S}|} & \mathbf{v}^{\rm{T}}_{\vec{S} \vec{\alpha}} \\
  \mathbf{u}_{\vec{\alpha} \vec{S}} & \mathbf{u}_{\vec{\alpha} \vec{B}} \mathbf{v}^{\rm{T}}_{\vec{B} \vec{\alpha}}  \end{array}
\right) \rm{.}
\end{align*}

\noindent In the first equality, we apply Eq.~(\ref{eq:modifiedCB4}) together with the fact that, for each term in the sum for which $\vec{L}$ contains any of the first $|\vec{S}|$ columns, the matrix in the first determinant factor of that term contains at least one column of all zeros, and so the term evaluates to zero. This brings the second line to a form which we recognize to be an expansion by complementary minors of the rows $\vec{S}^{\prime}$ in the larger matrix in the third line. Finally, we use $\mathbf{u}^{\prime}_{ \vec{\alpha}^{\prime} \vec{S}^{\prime}} = \mathbf{u}_{\vec{\alpha} \vec{S}}$ (and similarly for the other submatrices) to undo our initial row and column rearrangements and therefore obtain the formula, Eq.~(\ref{eq:modifiedCB}). $\square$ 

\section{Appendix D: Fixed-Parity Sum}
\label{sec:FPSum}

Here we prove

\setcounter{Hequation}{12}
\renewcommand{\theHequation}{S\arabic{Hequation}}

\begin{align}
&\sum_{\substack{\{\vec{\beta}_l \subset \vec{B}_l, \vec{\beta}_r \subset \vec{B}_r ||\vec{\beta}_l| + |\vec{\beta}_r| = |\vec{\alpha}| - |\vec{S}|\rm{,} \\ |\vec{\beta}_r| \ \mathrm{mod} \ 2 \ = \ p \}}} \det{\left[\mathbf{u}_{\vec{\alpha}, (\vec{\beta}_l, \vec{S}, \vec{\beta_r})} \right]} \det{\left[\mathbf{v}_{\vec{\alpha}, (\vec{\beta}_l, \vec{S}, \vec{\beta_r})} \right]} \nonumber \\ &\mbox{\hspace{15mm}} = \frac{(-1)^{|\vec{S}|}}{2} \left[\det\left(\begin{array}{c c}
  \mathbf{0}_{|\vec{S}| \times |\vec{S}|} & \mathbf{v}^{\rm{T}}_{\vec{S} \vec{\alpha}} \\
  \mathbf{u}_{\vec{\alpha} \vec{S}} & \mathbf{u}_{\vec{\alpha} \vec{B}} \mathbf{v}^{\rm{T}}_{\vec{B} \vec{\alpha}}  \end{array}
\right) + (-1)^p \det\left(\begin{array}{c c}
  \mathbf{0}_{|\vec{S}| \times |\vec{S}|} & \mathbf{{v}}^{\rm{T}}_{\vec{S} \vec{\alpha}} \\
  \mathbf{{u}}_{\vec{\alpha} \vec{S}} & \mathbf{{u}}_{\vec{\alpha} \vec{B}}\left(\mathbf{I} - 2 \mathbf{P}_{\vec{B}_r}\right)_{\vec{B} \vec{B}} \mathbf{{v}}^{\rm{T}}_{\vec{B} \vec{\alpha}}  \end{array}
\right) \right] \label{eq:parityCB} \rm{,}
\end{align}

\noindent where $\vec{S}$ is disjoint from $\vec{B} \equiv (\vec{B}_l, \vec{B}_r)$. The statement follows simply from

\begin{align}
&\left(\sum_{\substack{\{\vec{\beta}_l \subset \vec{B}_l, \vec{\beta}_r \subset \vec{B}_r ||\vec{\beta}_l| + |\vec{\beta}_r| = |\vec{\alpha}| - |\vec{S}|\rm{,} \\ |\vec{\beta}_r| \ \mathrm{even}\}}} - \sum_{\substack{\{\vec{\beta}_l \subset \vec{B}_l, \vec{\beta}_r \subset \vec{B}_r ||\vec{\beta}_l| + |\vec{\beta}_r| = |\vec{\alpha}| - |\vec{S}|\rm{,} \\ |\vec{\beta}_r| \ \mathrm{odd}\}}} \right) \det{\left[\mathbf{u}_{\vec{\alpha}, (\vec{\beta}_l, \vec{S}, \vec{\beta_r})} \right]} \det{\left[\mathbf{v}_{\vec{\alpha}, (\vec{\beta}_l, \vec{S}, \vec{\beta_r})} \right]} \nonumber \\ 
&\mbox{\hspace{10mm}} = \left[\sum_{\substack{\{\vec{\beta}_l \subset \vec{B}_l, \vec{\beta}_r \subset \vec{B}_r ||\vec{\beta}_l| + |\vec{\beta}_r| = |\vec{\alpha}| - |\vec{S}|\rm{,} \\ |\vec{\beta}_r| \ \mathrm{even}\}}} (-1)^{|\vec{\beta}_r|} + \sum_{\substack{\{\vec{\beta}_l \subset \vec{B}_l, \vec{\beta}_r \subset \vec{B}_r ||\vec{\beta}_l| + |\vec{\beta}_r| = |\vec{\alpha}| - |\vec{S}|\rm{,} \\ |\vec{\beta}_r| \ \mathrm{odd}\}}} (-1)^{|\vec{\beta}_r|} \right] \det{\left[\mathbf{u}_{\vec{\alpha}, (\vec{\beta}_l, \vec{S}, \vec{\beta_r})} \right]} \det{\left[\mathbf{v}_{\vec{\alpha}, (\vec{\beta}_l, \vec{S}, \vec{\beta_r})} \right]} \nonumber \\
&\mbox{\hspace{10mm}} = \sum_{\{(\vec{\beta}_l, \vec{\beta_r}) \subset \vec{B} | |\vec{\beta_l}| + |\vec{\beta_r}| = |\vec{\alpha}| - |\vec{S}|\}}  (-1)^{|\vec{\beta}_r|} \det{\left[\mathbf{u}_{\vec{\alpha}, (\vec{\beta}_l, \vec{S}, \vec{\beta_r})} \right]} \det{\left[\mathbf{v}_{\vec{\alpha}, (\vec{\beta}_l, \vec{S}, \vec{\beta_r})} \right]} \nonumber
\end{align}

\noindent In the first equality, we used the fact that $(-1)^{|\vec{\beta}_r|} = 1$ in the former sum, and $(-1)^{|\vec{\beta}_r|} = -1$ in the latter. In the second, we simply combined the sums over all even-sized $\vec{\beta}_r$ and all odd-sized $\vec{\beta}_r$ with the same summand into the sum over all $(\vec{\beta}_l, \vec{\beta}_r)$. We next apply the steps we used to prove the modified Cauchy-Binet formula, except now applying

\begin{align*}
\det\left[(\mathbf{I} - 2 \mathbf{P}_{\vec{B}_r})_{(\vec{\beta}_l, \vec{\beta}_r) (\vec{\beta}'_l, \vec{\beta}'_r)}\right] = (-1)^{|\vec{\beta}_r|} \delta_{\vec{\beta}_l \vec{\beta}'_l} \delta_{\vec{\beta}_r \vec{\beta}'_r}
\end{align*}

\noindent to evaluate the sum with the sign factor $(-1)^{|\vec{\beta}_r|}$ which appears in the place of Eq.~(\ref{eq:partialCB}). This gives

\begin{align}
& \left(\sum_{\substack{\{\vec{\beta}_l \subset \vec{B}_l, \vec{\beta}_r \subset \vec{B}_r ||\vec{\beta}_l| + |\vec{\beta}_r| = |\vec{\alpha}| - |\vec{S}|\rm{,} \\ |\vec{\beta}_r| \ \mathrm{even}\}}} - \sum_{\substack{\{\vec{\beta}_l \subset \vec{B}_l, \vec{\beta}_r \subset \vec{B}_r ||\vec{\beta}_l| + |\vec{\beta}_r| = |\vec{\alpha}| - |\vec{S}|\rm{,} \\ |\vec{\beta}_r| \ \mathrm{odd}\}}} \right) \det{\left[\mathbf{u}_{\vec{\alpha}, (\vec{\beta}_l, \vec{S}, \vec{\beta_r})} \right]} \det{\left[\mathbf{v}_{\vec{\alpha}, (\vec{\beta}_l, \vec{S}, \vec{\beta_r})} \right]} \nonumber \\
&\mbox{\hspace{15mm}} = (-1)^{|\vec{S}|} \det
\left(
\begin{array}{c c}
  \mathbf{0}_{|\vec{S}| \times |\vec{S}|} & \mathbf{v}^{\rm{T}}_{\vec{S} \vec{\alpha}} \\
  \mathbf{u}_{\vec{\alpha} \vec{S}} & \mathbf{u}_{\vec{\alpha} \vec{B}}(\mathbf{I} - 2 \mathbf{P}_{\vec{B}_r})_{\vec{B} \vec{B}} \mathbf{v}^{\rm{T}}_{\vec{B} \vec{\alpha}} \end{array}
\right) \nonumber \rm{,}
\end{align}

\noindent from Eq.~(\ref{eq:modifiedCB}). Applying this, together with

\setcounter{Hequation}{13}
\renewcommand{\theHequation}{S\arabic{Hequation}}

\begin{align}
& \left(\sum_{\substack{\{\vec{\beta}_l \subset \vec{B}_l, \vec{\beta}_r \subset \vec{B}_r ||\vec{\beta}_l| + |\vec{\beta}_r| = |\vec{\alpha}| - |\vec{S}|\rm{,} \\ |\vec{\beta}_r| \ \mathrm{even}\}}} + \sum_{\substack{\{\vec{\beta}_l \subset \vec{B}_l, \vec{\beta}_r \subset \vec{B}_r ||\vec{\beta}_l| + |\vec{\beta}_r| = |\vec{\alpha}| - |\vec{S}|\rm{,} \\ |\vec{\beta}_r| \ \mathrm{odd}\}}} \right) \det{\left[\mathbf{u}_{\vec{\alpha}, (\vec{\beta}_l, \vec{S}, \vec{\beta_r})} \right]} \det{\left[\mathbf{v}_{\vec{\alpha}, (\vec{\beta}_l, \vec{S}, \vec{\beta_r})} \right]} \nonumber \\
&\mbox{\hspace{15mm}} = (-1)^{|\vec{S}|} \det
\left(
\begin{array}{c c}
  \mathbf{0}_{|\vec{S}| \times |\vec{S}|} & \mathbf{v}^{\rm{T}}_{\vec{S} \vec{\alpha}} \\
  \mathbf{u}_{\vec{\alpha} \vec{S}} & \mathbf{u}_{\vec{\alpha} \vec{B}} \mathbf{v}^{\rm{T}}_{\vec{B} \vec{\alpha}}  \end{array}
\right) \nonumber
\end{align}

\noindent we solve for the sums over even $|\vec{\beta}_r|$ and odd $|\vec{\beta}_r|$ individually to obtain Eq.~(\ref{eq:parityCB}) above. $\square$

\section{Appendix E: Exact Calculation of the OTO Correlator -- Proof of Equation 6}
\label{sec:ExactOTO}

Here we give an explicit calculation of the matrix $\mathbf{M}_s$, defined implicitly by

\begin{align}
\frac{1}{2^{n + 2}}||[\mathbf{n}_s \cdot \bm{\sigma}_s, U^{\dagger} C_{\vec{\alpha}} U]||^2 \equiv \mathbf{n}^{*}_{s} \cdot \mathbf{M}_s \cdot \mathbf{n}_{s}
\label{eq:otomatrix}
\end{align}

\noindent for some single-spin operator $\mathbf{n}_s\cdot \bm{\sigma}_s$ acting on qubit $s$, matchgate unitary $U$, and Majorana configuration $C_{\vec{\alpha}}$, where $||A||^2 \equiv \tr{\left(A^{\dagger} A\right)}$. Since we are only considering a single spin $s$, we drop the spin labels on $\mathbf{n} \equiv \mathbf{n}_s$ and $\mathbf{M} \equiv \mathbf{M}_s$ in this section for convenience, choosing to label the components of these objects by subscripts instead. We begin by expanding the r.h.s., using this definition and Eq.~(\ref{eq:MMTransition})

\begin{align*}
&\frac{1}{2^{n + 2}}||[\mathbf{n} \cdot \bm{\sigma}_s, U^{\dagger} C_{\vec{\alpha}} U]||^2 \\ 
&\mbox{\hspace{10mm}}= \frac{1}{2^{n + 2}}\tr{\left\{\left[\sum_{k \in \{x, y, z \}} n_k \sigma^k_s, \sum_{\{\vec{\beta} | |\vec{\beta}| = |\vec{\alpha}| \}} \det{\left(\mathbf{u}_{\vec{\alpha} \vec{\beta}}\right)} C_{\vec{\beta}}\right]^{\dagger} \left[\sum_{k^{\prime} \in \{x, y, z \}} n_{k^{\prime}} \sigma^{k^{\prime}}_s, \sum_{\{\vec{\beta}^{\prime} | |\vec{\beta}^{\prime}| = |\vec{\alpha}| \}} \det{\left(\mathbf{u}_{\vec{\alpha} \vec{\beta}^{\prime}}\right)} C_{\vec{\beta}^{\prime}}\right] \right\}} \\
&\mbox{\hspace{10mm}}= \frac{1}{2^{n + 2}}\sum_{k, k^{\prime} \in \{x, y, z\}}n^{*}_k n_{k^{\prime}} \sum_{\substack{\{\vec{\beta} ||\vec{\beta}| = |\vec{\alpha}|\} \\ \{\vec{\beta}^{\prime} ||\vec{\beta}^{\prime}| = |\vec{\alpha}|\}}} \det{\left(\mathbf{u}_{\vec{\alpha} \vec{\beta}} \right)} \det{\left(\mathbf{u}_{\vec{\alpha} \vec{\beta}^{\prime}} \right)} \tr{\left\{\left[\sigma_s^k, C_{\vec{\beta}} \right]^{\dagger} \left[\sigma_s^{k^{\prime}}, C_{\vec{\beta}^{\prime}} \right] \right\}}
\end{align*}

\noindent Thus

\begin{align*}
M_{k k^{\prime}} \equiv \frac{1}{2^{n + 2}} \sum_{\substack{\{\vec{\beta} ||\vec{\beta}| = |\vec{\alpha}|\} \\ \{\vec{\beta}^{\prime} ||\vec{\beta}^{\prime}| = |\vec{\alpha}|\}}} \det{\left(\mathbf{u}_{\vec{\alpha} \vec{\beta}} \right)} \det{\left(\mathbf{u}_{\vec{\alpha} \vec{\beta}^{\prime}} \right)} \tr{\left\{\left[\sigma_s^k, C_{\vec{\beta}} \right]^{\dagger} \left[\sigma_s^{k^{\prime}}, C_{\vec{\beta}^{\prime}} \right] \right\}}
\end{align*}

\noindent We can obtain the diagonal elements of $\mathbf{M}$, for which $k = k^{\prime}$, from Theorem \ref{thm:efficientoto}.
It remains, then, to calculate the off-diagonal elements of $\mathbf{M}$. Clearly, $\mathbf{M} = \mathbf{M}^{\mathrm{T}}$ by the cyclic property of the trace and the fact that the sum over $\vec{\beta}$ and $\vec{\beta}^{\prime}$ is symmetric in these indices. Furthermore, we must have $M_{xz} = M_{yz} = 0$ since there are no $\vec{\beta}$ and $\vec{\beta}^{\prime}$ for which $|\vec{\beta}| = |\vec{\beta}^{\prime}|$ and which describe the same Pauli string on every spin except $s$, with a $X_s$ (or $Y_s$) present on one and a $Z_s$ present on the other. This is because the former are always described by exactly one Majorana operator and the latter by either two or zero Majorana operators. Therefore, we are only left to calculate $M_{xy}.$ In this case, we have

\begin{align*}
M_{xy} = \frac{1}{2^{n + 2}}\sum_{\substack{\{\vec{\beta} ||\vec{\beta}| = |\vec{\alpha}|\} \\ \{\vec{\beta}^{\prime} ||\vec{\beta}^{\prime}| = |\vec{\alpha}|\}}} \det{\left(\mathbf{u}_{\vec{\alpha} \vec{\beta}} \right)} \det{\left(\mathbf{u}_{\vec{\alpha} \vec{\beta}^{\prime}} \right)} \tr{\left\{\left[X_s, C_{\vec{\beta}} \right]^{\dagger} \left[Y_s, C_{\vec{\beta}^{\prime}} \right] \right\}}
\end{align*}

\noindent The only nonvanishing terms will be those for which $\vec{\beta}$ and $\vec{\beta}^{\prime}$ describe the same Pauli string on every spin except for $s$ and for which there is a $Y_s$ present for $C_{\vec{\beta}}$ and an $X_s$ present for $C_{\vec{\beta}^{\prime}}$. We examine the conditions under which an operator $\sigma^{k^{\prime}}_s$ will be present in $C_{\vec{\beta}}$. Let $\vec{\beta} = \left(\vec{\beta}_l, \vec{\beta}_s, \vec{\beta}_r \right)$, where $\vec{\beta}_l \subset [2(s - 1)]$ consists of all indices less than $2s - 1$, and $\vec{\beta}_r \subset \overline{[2s]}$ consists of all indices greater than $2s$ (corresponding to the spins to the left and right of $s$, respectively). We have

\setcounter{Hequation}{14}
\renewcommand{\theHequation}{S\arabic{Hequation}}

\begin{align}
X_s &\ \mathrm{present \ only \ if \ either: \ } \vec{\beta}_s =
\begin{cases}
\left(2s - 1\right) & \mathrm{and} \ |\vec{\beta}_r| \ \rm{even} \\
 \left(2s \right) & \mathrm{and} \ |\vec{\beta}_r| \ \rm{odd} \\
\end{cases} \nonumber \\
Y_s &\ \mathrm{present \ only \ if \ either: \ } \vec{\beta}_s =
\begin{cases}
\left(2s\right) & \mathrm{and} \ |\vec{\beta}_r| \ \rm{even} \\
 \left(2s - 1\right) & \mathrm{and} \ |\vec{\beta}_r| \ \rm{odd} \\
\end{cases} \label{eq:pauliconds} \\
Z_s &\ \mathrm{present \ only \ if \ either: \ } \vec{\beta}_s =
\begin{cases}
\left(2s - 1, 2s\right) & \mathrm{and} \ |\vec{\beta}_r| \ \rm{even} \\
 \left( \right) & \mathrm{and} \ |\vec{\beta}_r| \ \rm{odd} \\
\end{cases} \nonumber
\end{align}

\noindent When $|\vec{\beta}_r|$ is odd, then $C_{\vec{\beta}}$ and $C_{\vec{\beta}^{\prime}}$ will have a relative phase of $-1$ between them since $X_s Z_s = -i Y_s$, but $Y_s Z_s = i X_s$. Additionally taking into account the fact that $[X_s, Y_s] = -[Y_s, X_s]$ gives

\begin{align*}
M_{xy} &= \left(\sum_{\substack{\{(\vec{\beta}_l, \vec{\beta}_r) | |\vec{\beta}_l| + |\vec{\beta}_r| + 1 = |\vec{\alpha}|\rm{,} \\ |\vec{\beta}_r| \ \rm{odd} \}}} - \sum_{\substack{\{(\vec{\beta}_l, \vec{\beta}_r) | |\vec{\beta}_l| + |\vec{\beta}_r| + 1 = |\vec{\alpha}|\rm{,} \\ |\vec{\beta}_r| \ \rm{even} \}}} \right) \det{\left[\mathbf{u}_{\vec{\alpha} (\vec{\beta}_l, 2s - 1, \vec{\beta}_r)}\right]} \det{\left[\mathbf{u}_{\vec{\alpha} (\vec{\beta}_l, 2s, \vec{\beta}_r)}\right]} \\
M_{xy} &= \det{\left(
\begin{array}{c c}
0 & \mathbf{u}^{\mathrm{T}}_{(2s) \vec{\alpha}} \\
\mathbf{u}_{\vec{\alpha} (2s - 1)} & \mathbf{u}_{\vec{\alpha} \overline{(2s - 1, 2s)}} \left(\mathbf{I} - 2 \mathbf{P}_{\overline{[2s]}}\right)_{\overline{(2s - 1, 2s)} \overline{(2s - 1, 2s)}} \mathbf{u}^{\mathrm{T}}_{\overline{(2s - 1, 2s)} \vec{\alpha}}
\end{array}\right)} \rm{,}
\end{align*}

\noindent therefore completing the calculation. For completeness, we list the elements of $\mathbf{M}$, once again, below

\begin{align*}
M_{xx} &= \frac{1}{2} \left\{1 - (-1)^{|\vec{\alpha}|} \det [\mathbf{u}_{\vec{\alpha} [2n]} (\mathbf{I} - 2\mathbf{P}_{[2s - 1]}) \mathbf{u}^{\mathrm{T}}_{[2n] \vec{\alpha}}] \right\} \\
M_{yy} &= \frac{1}{2} \left\{1 - (-1)^{|\vec{\alpha}|} \det [\mathbf{u}_{\vec{\alpha} [2n]} (\mathbf{I} - 2\mathbf{P}_{[2(s - 1)] \cup (2s)}) \mathbf{u}^{\mathrm{T}}_{[2n] \vec{\alpha}}] \right\} \\
M_{zz} &= \frac{1}{2} \left\{1 - \det [\mathbf{u}_{\vec{\alpha} [2n]} (\mathbf{I} - 2\mathbf{P}_{(2s - 1, 2s)}) \mathbf{u}^{\mathrm{T}}_{[2n] \vec{\alpha}}] \right\} \\
M_{xy} = M_{yx} &=  \det{\left(
\begin{array}{c c}
0 & \mathbf{u}^{\mathrm{T}}_{(2s) \vec{\alpha}} \\
\mathbf{u}_{\vec{\alpha} (2s - 1)} & \mathbf{u}_{\vec{\alpha} \overline{(2s - 1, 2s)}} \left(\mathbf{I} - 2 \mathbf{P}_{\overline{[2s]}}\right)_{\overline{(2s - 1, 2s)} \overline{(2s - 1, 2s)}} \mathbf{u}^{\mathrm{T}}_{\overline{(2s - 1, 2s)} \vec{\alpha}}
\end{array}\right)} \\
M_{xz} = M_{zx} = M_{yz} = M_{zy} &= 0 \mbox{\hspace{80mm}} \square
\end{align*} 

\section{Appendix F: Proof of Theorem 2}
\label{sec:thm2proof}

Here we prove the statement

\setcounter{Hequation}{15}
\renewcommand{\theHequation}{S\arabic{Hequation}}

\begin{align}
\overline{|\langle O \rangle - \langle (\otimes_{s \in S} \mathcal{E}_s)(O) \rangle|} \leq \frac{1}{\sqrt{2^{n + 2}}} \sum_{s \in S} ||[\mathbf{n}_s\cdot \bm{\sigma}_s, O]|| \equiv \sum_{s \in S} \sqrt{\mathbf{n}^{*}_{s} \cdot \mathbf{M}_s \cdot \mathbf{n}_s} \label{eq:Theorem1} \rm{,}
\end{align} 

\noindent where the average $\overline{(\cdot)}$ is taken over a product basis $\{\otimes_{j = 1}^n |(-1)^{\ell_j} \mathbf{n}^{\bot}_j \rangle \}_{\bm{\ell} \in \{0, 1\}^{\times n}}$ whose Bloch axes are orthogonal to the $\{\mathbf{n}_s\}_s$, and $\mathcal{E}_s$ is the depolarizing channel on qubit $s$, given by

\setcounter{Hequation}{16}
\renewcommand{\theHequation}{S\arabic{Hequation}}

\begin{align}
\mathcal{E}_s(O) = \frac{1}{4}\left(O + \sum_{k \in \{x, y, z\}}  \sigma_s^k O \sigma_s^k \right)\rm{.}
\label{eq:depolardef}
\end{align}

\noindent This channel has the effect of projecting $O$ onto its component which acts only as the identity on site $s$. We further assume that $O$ is a traceless Hermitian operator. We first show that the left-hand side of Eq.~(\ref{eq:Theorem1}) is bounded by the trace norm for any averaging orthonormal basis set $\{|j\rangle\}_j$ and Hermitian operator $A$, as
	
\setcounter{Hequation}{17}
\renewcommand{\theHequation}{S\arabic{Hequation}}

\begin{align}
\overline{|\langle A\rangle|} &\equiv \frac{1}{2^n} \sum_{j = 1}^{2^n} |\langle j | A |j \rangle| \nonumber \\
&= \frac{1}{2^n} \sum_{j = 1}^{2^n} | \sum_{k = 1}^{2^n} U^{*}_{kj} d_k U_{kj} | \nonumber \\
&\leq \frac{1}{2^n} \sum_{j, k = 1}^{2^n} |d_k| |U_{kj}|^2 \nonumber \\
&= \frac{1}{2^n} \sum_{k = 1}^{2^n} |d_k| \nonumber \\
\overline{|\langle A \rangle|} &\leq \frac{1}{2^n} \tr{|A|} \label{eq:trdistance}
\end{align}

\noindent From the first to the second line, we used the fact that we assume $A$ to be a Hermitian operator, and so can be diagonalized as $A = U^{\dagger} D U$, for $D$ a diagonal matrix. From the second to the third line, we applied the triangle inequality, and from the third to the fourth line, we rearranged sums and used the fact that $\sum_{j = 1}^{2^n} |U_{kj}|^2 = 1$. The resulting quantity is the trace norm.

Without loss of generality, let $S = \{1, 2, \dots, |S|\}$. Then we have, for any input state

\setcounter{Hequation}{18}
\renewcommand{\theHequation}{S\arabic{Hequation}}

\begin{align}
|\langle O \rangle - \langle (\otimes_{s \in S} \mathcal{E}_s)(O) \rangle| &= |\langle O \rangle + \sum_{k = 1}^{|S| - 1} \left[ \langle \left(\otimes_{j = 1}^k \mathcal{E}_j \right)(O)\rangle - \langle \left(\otimes_{j = 1}^k \mathcal{E}_j \right)(O)\rangle \right] - \langle (\otimes_{s \in S} \mathcal{E}_s)(O) \rangle| \nonumber \\
&= |\sum_{k = 1}^{|S|} \left\{ \langle \left(\otimes_{j = 1}^{k - 1} \mathcal{E}_j \right)(O) \rangle -  \langle \left(\otimes_{j = 1}^{k - 1} \mathcal{E}_j \right)\left[\mathcal{E}_k (O) \right]\rangle\right\}| \nonumber \\
|\langle O \rangle - \langle (\otimes_{j \in S} \mathcal{E}_j)(O) \rangle| &\leq \sum_{k = 1}^{|S|} | \langle \left(\otimes_{j = 1}^{k - 1} \mathcal{E}_j \right)(O) \rangle -  \langle \left(\otimes_{j = 1}^{k - 1} \mathcal{E}_j \right)\left[\mathcal{E}_k (O) \right]\rangle| \rm{,}
\label{eq:telescopesum}
\end{align}

\noindent where, from the first to the second line, we expressed the difference as a telescoping sum (let $\otimes_{j = 1}^{0}\mathcal{E}_j(O) \equiv O$) and applied the triangle inequality in the third line. Next, we use the fact that, for any single-spin input $|\mathbf{n}^{\bot}\rangle$, the depolarized operator expectation value is the same as that of the dephased operator in any basis orthogonal to the input. That is, let $\left(\mathbf{n}\cdot \bm{\sigma}\right)|\mathbf{n}^{\bot}\rangle = e^{i\phi}|-\mathbf{n}^{\bot}\rangle$ for some phase $\phi$, and we have

\setcounter{Hequation}{19}
\renewcommand{\theHequation}{S\arabic{Hequation}}

\begin{align}
|\langle O \rangle - \langle \mathcal{E}_s(O) \rangle| = |\langle O \rangle - \langle \frac{1}{2}\left[O + \left(\mathbf{n}\cdot \bm{\sigma}_s\right) O \left(\mathbf{n}\cdot \bm{\sigma}_s\right) \right] \rangle| = \frac{1}{2} |\langle O\rangle - \langle \left(\mathbf{n}\cdot \bm{\sigma}_s\right) O \left(\mathbf{n}\cdot \bm{\sigma}_s\right) \rangle | \label{eq:dephase}
\end{align}

\noindent on input state $|\mathbf{n}^{\bot}\rangle$ on qubit $s$. Furthermore, taking the same quantity for input state $|-\mathbf{n}^{\bot}\rangle$ has the effect of exchanging $O$ and $\left(\mathbf{n}\cdot \bm{\sigma}_s\right) O \left(\mathbf{n}\cdot \bm{\sigma}_s\right)$ inside the absolute value, leaving it unchanged. Thus, we average over a product basis of $\{\otimes_{j = 1}^n |(-1)^{\ell_j} \mathbf{n}^{\bot}_j \rangle \}_{\bm{\ell} \in \{0, 1\}^{\times n}}$ on both sides of Eq.~(\ref{eq:telescopesum}) and apply Eq.~(\ref{eq:dephase}) to obtain 

\begin{align*}
\overline{| \langle O \rangle - \langle (\otimes_{s \in S} \mathcal{E}_s)(O) \rangle|} &\leq \frac{1}{2^{n + 1}}\sum_{k = 1}^{|S|} \tr | \left(\otimes_{j = 1}^{k - 1} \mathcal{E}_j \right)(O) - \left(\otimes_{j = 1}^{k - 1} \mathcal{E}_j \right)\left[\left(\mathbf{n}_{k}\cdot \bm{\sigma}_k\right) O \left(\mathbf{n}_{k}\cdot \bm{\sigma}_k\right)\right]| \\
\overline{| \langle O \rangle - \langle (\otimes_{s \in S} \mathcal{E}_s)(O) \rangle|} &\leq \frac{1}{2^{n + 1}} \sum_{k = 1}^{|S|} \tr | O - \left(\mathbf{n}_{k}\cdot \bm{\sigma}_k\right) O \left(\mathbf{n}_{k}\cdot \bm{\sigma}_k\right) |\rm{,}
\end{align*}

\noindent using the bound (\ref{eq:trdistance}) in the second line and the fact that the depolarizing channel cannot increase the trace distance between operators in the third. Finally, we apply the operator norm inequality

\begin{align*}
\tr|A| \equiv \tr \sqrt{A^{\dagger} A} \leq \sqrt{r \tr A^{\dagger}A} \leq \sqrt{2^n \tr A^{\dagger}A} \rm{,}
\end{align*}

\noindent where $r \leq 2^n$ is the rank of $A$, together with the fact that 

\begin{align*}
\tr\left\{\left[O - \left(\mathbf{n}_{k}\cdot \bm{\sigma}_k\right) O \left(\mathbf{n}_{k}\cdot \bm{\sigma}_k\right)\right]^{\dagger} \left[O - \left(\mathbf{n}_{k}\cdot \bm{\sigma}_k\right) O \left(\mathbf{n}_{k}\cdot \bm{\sigma}_k \right)\right] \right\} = ||[\mathbf{n}_{k}\cdot \bm{\sigma}_k, O]||^2 \rm{,}
\end{align*}

\noindent for $||\cdot||$ the Frobenius norm, to arrive at our result 

\begin{align*}
\overline{| \langle O \rangle - \langle (\otimes_{s \in S} \mathcal{E}_s)(O) \rangle|} &\leq \frac{1}{2^{n + 1}} \sum_{s \in S} \tr | O - \left(\mathbf{n}_s\cdot \bm{\sigma}_s\right) O \left(\mathbf{n}_s\cdot \bm{\sigma}_s\right) | \\ &\leq \frac{\sqrt{2^n}}{2^{n + 1}} \sum_{s \in S} ||[\mathbf{n}_s\cdot \bm{\sigma}_s, O]|| \\
\overline{| \langle O \rangle - \langle (\otimes_{s \in S} \mathcal{E}_s)(O) \rangle|} &\leq \frac{1}{\sqrt{2^{n + 2}}} \sum_{s \in S} ||[\mathbf{n}_s\cdot \bm{\sigma}_s, O]|| \equiv \sum_{s \in S} \sqrt{\mathbf{n}^{*}_{s} \cdot \mathbf{M}_s \cdot \mathbf{n}_s} \rm{.} \mbox{\hspace{10mm}} \square
\end{align*} 

\noindent We remark that the bound is a sum of terms which are always between zero and one, and by Markov's inequality, we can use this result to bound the fraction of these basis states for which the change in expectation value exceeds some threshold $\delta$

\begin{align*}
\mu \left[ |\langle O \rangle - \langle (\otimes_{s \in S} \mathcal{E}_s)(O) \rangle| \geq \delta \right] \leq \frac{1}{\delta} \sum_{s \in S} \sqrt{\mathbf{n}^{*}_{s} \cdot \mathbf{M}_s \cdot \mathbf{n}_s} \rm{.}
\end{align*}

\noindent Since we expect each of these terms to be exponentially decaying outside of the lightcone, depolarizing in this region should only induce a change in expectation value greater than the threshold for some exponentially small set of input states. This affords us exponential precision so long as we are willing to tolerate a fixed fraction of these ``pathological" states. Finally, since we had a freedom in choosing which basis over which to depolarize in Eq.~(\ref{eq:dephase}), we may optimize each of the $\{\mathbf{n}_s\}_s$ over the subspace for which they are orthogonal to the Bloch axes of the input to make this bound as restrictive as possible.

\section{Appendix G: Numerical Analysis}
\label{sec:numanalysis}

We characterize the propagation of $B(t)$ by taking the singular value decomposition in discrete time

\setcounter{Hequation}{20}
\renewcommand{\theHequation}{S\arabic{Hequation}}

\begin{align}
||[\mathbf{n}_s \cdot \bm{\sigma}_s, B(t_k)]|| = \sum_j \lambda_j u_{j}(t_k) v_{j}(s) \rm{,}
\end{align}

\noindent where $t_k \equiv k \delta t$, and $\lambda_1 > \lambda_2 > \dots$ for the singular values $\lambda_j$. We take the function $u_{1}(t)$ to represent the light cone envelope and $v_{1}(s)$ the decay profile of $||[\mathbf{n}_s \cdot \bm{\sigma}_s, B(t)]||$ outside of this envelope. This method has several advantages: (i) the principal singular value component is the closest \emph{product} approximation to the light cone, and so has the form of the right-hand-side of Eq.~(\ref{eq:ZVLRBound}) in the main text; (ii) the singular values themselves give the error incurred in the approximation; and (iii) the principal singular value component is robust to fluctuations from specific disorder realizations, greatly reducing the number of samples needed. It therefore gives an operationally meaningful, robust, and numerically inexpensive means of extracting the envelope and decay profile, which is completely general beyond the setting of matchgate circuits considered here.

\begin{figure}
\includegraphics[width=0.95\columnwidth]{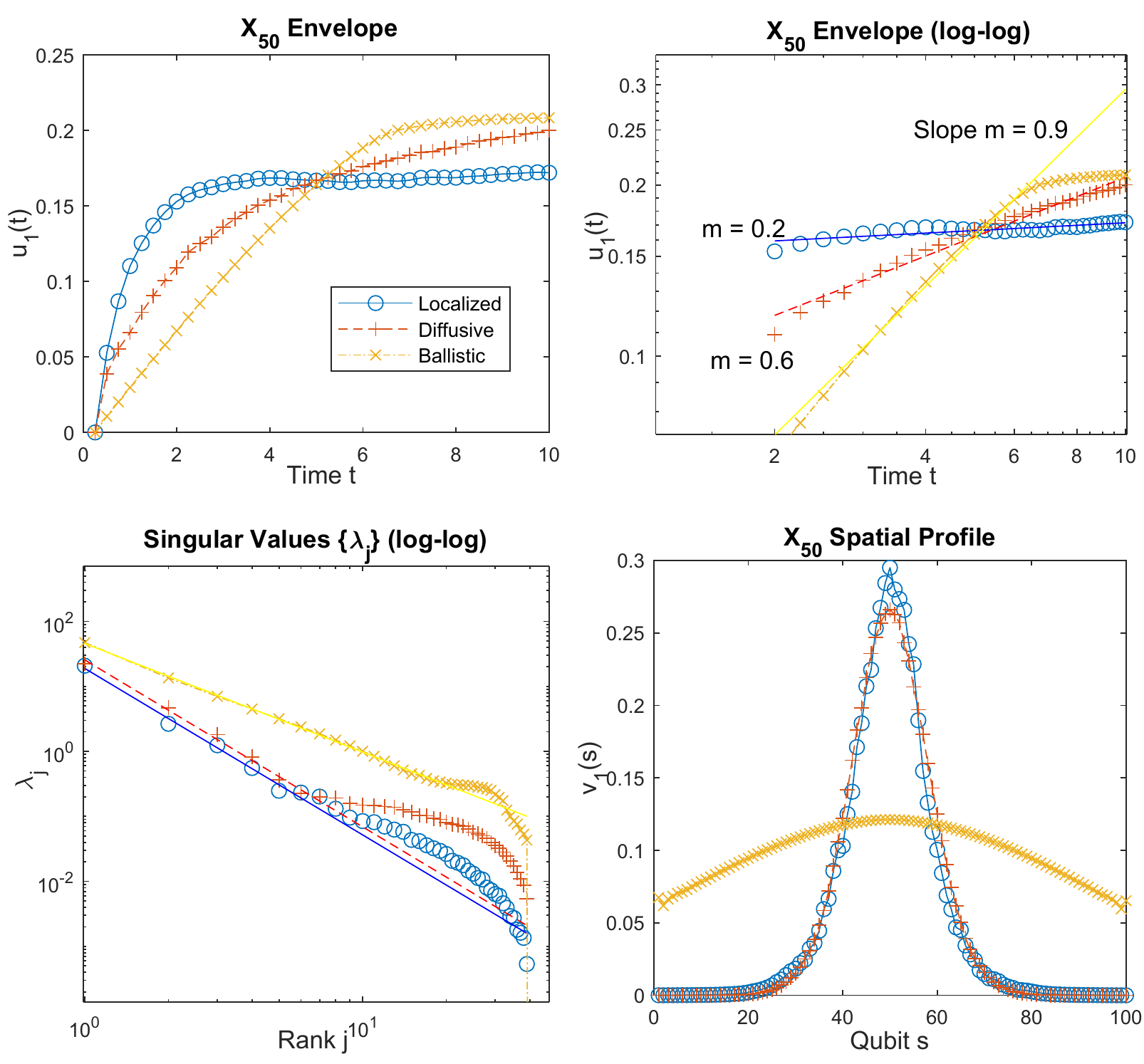}
\caption{(Color online) (Top, left) Envelopes of the $X$ light cones in Fig.~\ref{fig:propagation}, given by the principal temporal components of their singular value decompositions. (Top, right) Envelopes on a log-log plot, with fitted slopes displayed. Note the saturation of the ballistic case around $t \simeq 6$ is a boundary effect; otherwise, it is expected to be a straight slope. (Bottom, left) Singular values in these decompositions on log-log scale, which demonstrate the error in truncating to a product function; we see that these decay by several orders of magnitude over the first ten. (Bottom, right) Light cone decay profiles, given by the principal spatial vectors; these profiles are very nearly Gaussian, rather than exponential.}
\label{fig:svdprincipal}
\end{figure}

In Fig.~\ref{fig:svdprincipal}, we plot the results of our analysis for the $X$ light cones of Fig.~\ref{fig:propagation} in the main text. We see that the envelopes (linear scale on top left, and log-log scale on top right) $u_1(t)$ propagate as polynomials with different exponents $m$, which we take to be indicative of the dynamical phases of these profiles.

\end{document}